\documentclass[aip, pof,graphicx]{revtex4-2}

\usepackage{graphicx}% Include figure files
\usepackage{dcolumn}% Align table columns on decimal point
\usepackage{bm}% bold math
\usepackage[utf8]{inputenc}
\usepackage[T1]{fontenc}
\usepackage{amssymb}
\usepackage{mathptmx}
\usepackage{color}
\usepackage{longtable}
\usepackage{epstopdf}%This line makes .eps figures into .pdf - please comment out if not required.
\usepackage{multirow}

\draft % marks overfull lines with a black rule on the right

\begin{document}

%\preprint{}

\title[Bidisperse micro fluidized beds: Effect of bed inclination on mixing]{Bidisperse micro fluidized beds: Effect of bed inclination on mixing\\
	This article appeared in Phys. Fluids 36, 013303 (2024) and may be found at https://doi.org/10.1063/5.0179153.} %Title of paper



\author{Henrique B. Oliveira}
\affiliation{Faculdade de Engenharia Mec\^anica, Universidade Estadual de Campinas (UNICAMP),\\
	Rua Mendeleyev, 200, Campinas, SP, Brazil
}

\author{Erick M. Franklin}%
\email{erick.franklin@unicamp.br}
\thanks{Corresponding author}
\affiliation{Faculdade de Engenharia Mec\^anica, Universidade Estadual de Campinas (UNICAMP),\\
	Rua Mendeleyev, 200, Campinas, SP, Brazil
}


\date{\today}

\begin{abstract}	
Micro fluidized beds are basically suspensions of solid particles by an ascending fluid in a mm-scale tube, with applications in chemical and pharmaceutical processes involving powders. Although in many applications beds are polydisperse, previous works considered only monodisperse beds aligned in the vertical direction. However, introducing an inclination with respect to gravity leads to different bed patterns and mixing levels, which can be beneficial for some applications. In this paper, we investigate experimentally the behavior of micro gas-solid beds consisting of bidisperse mixtures under different inclinations. In our experiments, mono and bidisperse beds are filmed with a high-speed camera and the images are processed for obtaining measurements at both the bed and grain scales. We show that the degree of segregation is larger for vertical beds, but mixing varies non-monotonically with inclination, with an optimal angle of 30$^{\circ}$--50$^{\circ}$ with respect to gravity. By computing the mean and fluctuation velocities of grains, we reveal that the mixing layer results from the competition between segregation by kinetic sieving and circulation promoted by the fluid flow. We also observe worse fluidization as the angle relative to gravity increases, accounting then for the non-monotonic behavior. Our results bring new insights into mixing and segregation in polydisperse beds, which can be explored for processing powders in industry.
\end{abstract}


\maketitle %\maketitle must follow title, authors, abstract and \pacs

\section{INTRODUCTION}
\label{sec:intro}

A fluidized bed consists in the suspension of solid particles by a fluid flowing upwards, the grains' weight being balanced by forces exerted by the fluid. Because of their simple construction and high rates of mass and heat transfers, beds fluidized in large tubes (with diameter $D$ much higher than those of grains $d$) are frequently found in industry. In those cases, the ensemble of solids seems to behave as a fluid (although this is not really the case), which explains the term \textit{fluidized}. The apparent fluid-like behavior, however, is not observed when the bed is not large enough (when $D$ and $d$ have a difference of only one order of magnitude at most).

Although there is not a formal classification, former works \cite{Cunez, Cunez2, Cunez3, Cunez4} differentiate between narrow beds, for which 10 $<$ $D/d$ $\lesssim$ 100, and very-narrow beds, for which $D/d$ $\leq$ 10. In both cases, the structures appearing in the bed are different with respect to large beds, with transverse waves, blobs and bubbles appearing for narrow beds \cite{Duru,Duru2} and plugs and large bubbles for very-narrow beds \cite{Cunez, Cunez2}. In particular, the fluid-like behavior of grains is completely lost in the very-narrow case, with even clogging, crystallization (faint oscillations at the grain scale only) and jamming (absence of motion even at the grain scale) stopping fluidization \cite{Cunez3, Cunez4, Oliveira}.

C\'u\~nez and Franklin \cite{Cunez, Cunez2} investigated the dynamics of individual grains and structures appearing in very-narrow solid-liquid fluidized beds (SLFBs) for both mono \cite{Cunez} and bidisperse \cite{Cunez2} beds. C\'u\~nez and Franklin \cite{Cunez} found that instabilities in the form of alternating high- and low-compactness regions, called plugs and bubbles, respectively, appear in the bed due to dense networks of contact forces that percolate within the bed until reaching the tube wall (due to confinement). High confinement affects also the segregation and layer inversion in bidisperse very-narrow SLFBs, as showed by C\'u\~nez and Franklin \cite{Cunez2}, presenting a distinct behavior with respect to large beds. Later, C\'u\~nez and Franklin \cite{Cunez3} proceeded as Goldman and Swinney \cite{Goldman} and investigated very-narrow SLFBs under partial de-fluidization and re-fluidization. However, different from Goldman and Swinney \cite{Goldman}, their beds had $D/d$ $\leq$ 4.2 (high confinement) and different grain types were tested. They found that crystallization can occur at fluid velocities above that for minimum fluidization ($U_{mf}$), that different lattices can appear, that in the very-narrow case crystallization does not depend on the deceleration rate of the fluid flow, and that the jamming intensity depends on the particle type. Still for very-narrow SLFBs, Oliveira et al. \cite{Oliveira} showed that crystallization and refluidization can alternate successively along time in monodisperse beds of regular spheres, and found the characteristic times for crystallization. Interestingly, they showed that crystallization can be avoided by placing a layer of less-regular spheres on the bottom of the regular ones (those that crystallize when alone in the tube), mitigating the problem (since crystallization is often undesirable in fluidized beds). Finally, they showed that the high levels of agitation within the bottom layer hinders crystallization of the top layer.

Micro fluidized beds (MFBs) are those taking place in a mm- or cm-scale tube \cite{Zhang, Qie}, with applications in chemical and pharmaceutical processes involving powders, such as particle encapsulation \cite{Schreiber, Rodriguez}, pyrolysis \cite{Jia, Gao, Mao, Yu}, catalytic cracking \cite{Boffito, Guo2}, gasification \cite{Zeng, Zhang2, Cortazar}, capture of $CO_2$ \cite{Fang, Shen}, bioproduction \cite{Wu, Liu4, Pereiro}, and wastewater treatment \cite{Kuyukina, Kwak}. Because of the small diameter of the tube, MFBs are usually of the very-narrow type (the grains' diameter being of the order of hundreds of microns). However, given the grain scale and area of contact with the tube wall, adhesion forces are significant, and, therefore, structures and grain motion in MFBs are distinct from those in larger very-narrow beds.

The great potential of using MFBs in mechanical and chemical processes, in particular in the pharmaceutical industry for producing tablets and pills \cite{Kornblum, Katstra, Fung, Ervasti, Azad, Azad2, Shi} and vaccines in powder form \cite{Jiang, Huang, Gomez, Heida, Amorij}, has motivated recent investigations on MFBs. Still, most of those studies concerned cm-scale beds only, and, to the best of the authors' knowledge, all of them investigated vertical beds (without considerable inclinations with respect to gravity). For example, Guo et al. \cite{Guo} carried out experiments in gas-solid MFBs for tubes of varying diameters (from 4.3 to 25.5 mm), and found that classical correlations for the minimum fluidization velocity $U_{mf}$ do not work for MFBs: $U_{mf}$ is relatively much higher than in large beds. An  explanation for higher $U_{mf}$ in MFBs was proposed later by Do Nascimento et al. \cite{Nascimento} based on experiments in liquid-solid MFBs with hydraulic diameters of 1 and 2 mm: they found that adhesion forces in MFBs can reach values comparable to the hydrodynamic and gravitational ones. Those results were corroborated by other studies \cite{Li2, Rao, Wang3, Doroodchi}, all of them carrying out measurements at the bed scale only. More recently, C\'u\~nez and Franklin \cite{Cunez5} investigated experimentally a gas-solid mm-scale MFB with $D/d$ = 6 (in the very-narrow case) by carrying out measurements at both the bed and grain scales. They found that mm-scale beds have much reduced agitation (and, thus, transfer rates) when compared to dm- and m-scale beds, and that increasing the flow velocity increases only slightly the agitation of grains (not avoiding the appearance of plugs).

Although previous studies shed light on the physics of MFBs, none of them investigated the behavior of bidisperse beds nor the effects of bed inclination on fluidization. Many questions remain, thus, to be investigated. For instance, how grains segregate in a MFB? How bed inclination affects segregation? Is mixing improved by inclining the bed? What is the limit angle for fluidizing grains in MFBs? In this paper, we inquire into these questions by carrying out experiments with gas-solid MFBs of bidisperse mixtures under different inclinations. In our experiments, the tube diameter was $D$ = 3 mm and the bed was either mono or bidisperse, with mean particle diameter $d$ = 0.5 mm in both cases ($D/d$ = 6, corresponding to the very-narrow case). The bed was filmed with a high-speed camera and the images were processed for obtaining measurements at both the bed and grain scales. We show that the degree of segregation is larger for vertical beds, and that mixing varies non-monotonically with inclination, with an optimal angle for mixing of 30$^{\circ}$--50$^{\circ}$ with respect to gravity. By computing the mean and fluctuation velocities of grains, we show that the mixing layer results from the competition between segregation by kinetic sieving and circulation promoted by the fluid flow. We also observe worse fluidization as the angle relative to gravity increases, accounting then for the non-monotonic behavior, and present a diagram of mixing vs. deviation angle that can be useful for determining optimal mixing. Our results represent a new step for understanding mixing and segregation in MFBs, with potential applications in mechanical, chemical and pharmaceutical processes.

\section{EXPERIMENTAL SETUP}
\label{sec:exp_setup}

\begin{figure}[ht]
\centering
	\includegraphics[width=0.7\columnwidth]{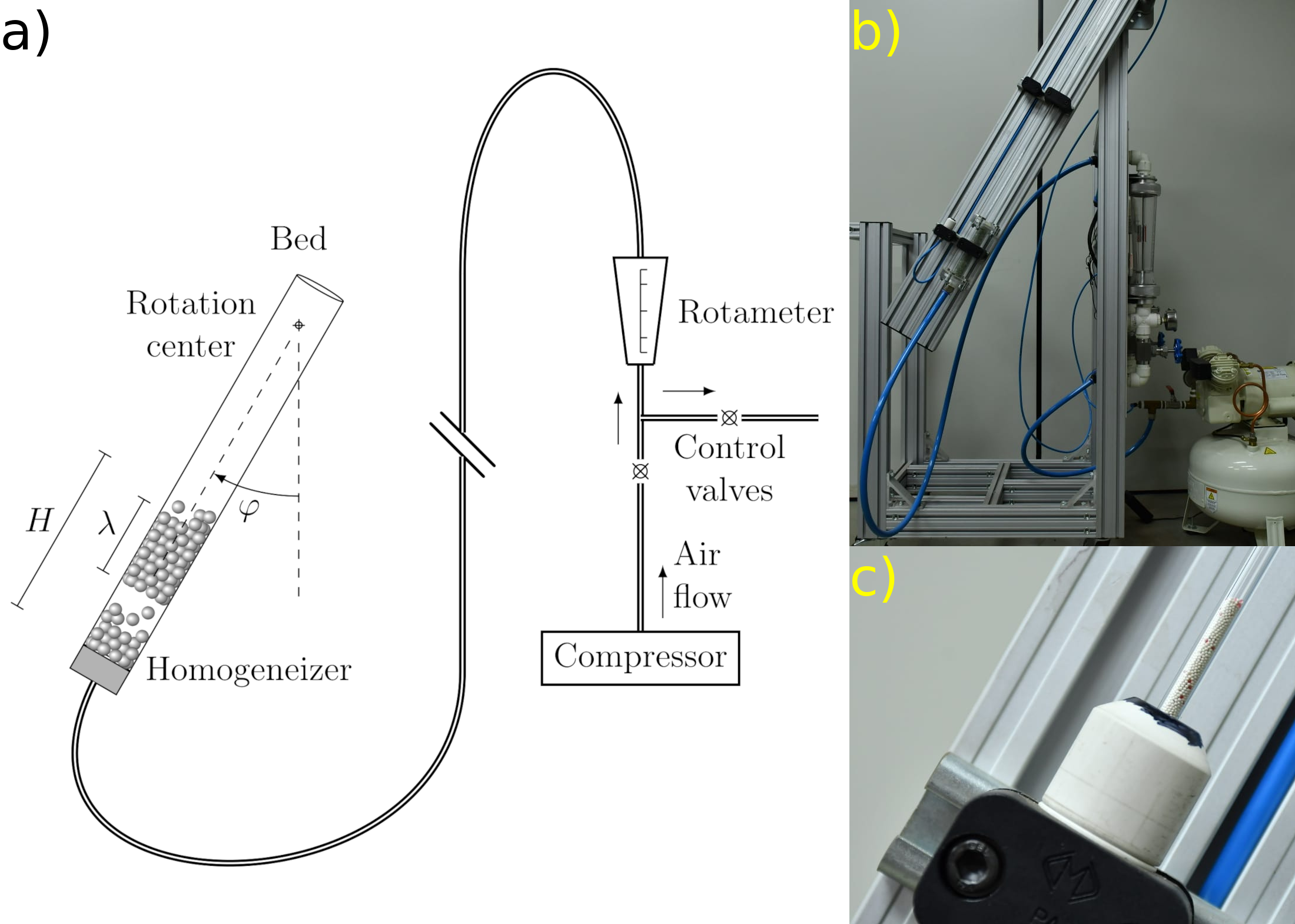}
	\caption{(a) Layout and (b) photograph of the experimental setup. (c) Photograph of part of the test section with the grains inside.}
	\label{fig_layout}
\end{figure} 

The experimental setup consisted of an air compressor, a pressure vessel, controlling vanes, termocouples, pressure transducers,  flow meters, a flow homogenizer, and a 3-mm-ID tube that can pivot in the vertical plane and be locked in predefined inclinations within 0$^\circ$ and 90$^\circ$ $\pm$ 0.5$^\circ$. Figure \ref{fig_layout} shows a layout and a photograph of the experimental setup, and more details of the pivoting system are available in the supplementary material.

We used glass (density $\rho_{p,1}$ = 2500 kg/m$^3$) and zirconium ($\rho_{p,2}$ = 4100 kg/m$^3$) spheres for the grains, which we call in the following species 1 and 2, respectively, both with diameters ($d_1$ and $d_2$) within 0.4 and 0.6 mm ($D/d$ = 6), and the fluid was air (microscopy images of the used spheres are available in the supplementary material). For the different conditions tested, we weighted the grains until reaching the desired masses of glass ($m_1$) and zirconium ($m_2$), and the total mass $m_T$ = $m_1$ $+$ $m_2$ was inserted in the test section (settled by gravity). Prior to each run, the tube was set in the vertical position and the bed fluidized by imposing a pre-determined air flow. In the bidisperse case, the bed was fluidized until steady segregation was observed (all glass spheres forming a top layer over the zirconium spheres). Afterward, the flow was stopped and the bed put in the desired angle $\varphi$ with respect to gravity. The test run then started by imposing the desired air velocity and the bed was filmed for 30 s.

In the experiments, the total mass of particles $m_T$ varied within 0.3 and 1.0 g, the initial heights of beds (when in the vertical position) $h_{if}$ within 24 and 60 mm, the cross-sectional mean velocities of air $\overline{U}$ within 1.57 and 2.16 m/s, and the bed inclination $\varphi$ (with respect to gravity) within 0 and 60$^\circ$. With that, the numbers of Stokes $St_t \,=\, v_t d \rho_p / (9\mu_f)$ and Reynolds $Re_t \,=\, \rho_f  v_t d / \mu_f$ based on the terminal velocity of one single particle $v_t$ are $St_t$  = 2.86 $\times$ 10$^4$ and 6.44 $\times$ 10$^4$ for species 1 and 2, respectively, and $Re_t$ = 1.26 $\times$ 10$^2$ and 1.73 $\times$ 10$^2$ for species 1 and 2, respectively, $\mu_f$ being the dynamic viscosity of the fluid. The settling velocity estimated using the Richardson--Zaki correlation, $v_s = v_t \left( 1-\phi_0 \right) ^{2.4}$, is 0.70  and 0.96 m/s for species 1 and 2, respectively, in which we considered $\phi_0$ $\approx$ 0.5 (based on C\'u\~nez and Franklin \cite{Cunez3}). The tested conditions are listed in Tab. \ref{tab_exp}, which presents also the Reynolds numbers for both tube and grains based on the cross-sectional mean velocity, $Re_D \,=\, \rho_f  \overline{U} D / \mu_f$ and $Re_d \,=\, \rho_f  \overline{U} d / \mu_f$, respectively.

\begin{table}
	\begin{center}
		\caption{Summary of tested conditions: mass of each species $m_1$ and $m_2$, fluid velocity for incipient fluidization $U_{if}$, bed height at incipient fluidization $h_{if}$, cross-sectional mean velocity normalized by that of incipient fluidization $\overline{U}/U_{if}$, and Reynolds numbers for the tube and grains based on the cross-sectional mean velocity, $Re_D$ and $Re_d$, respectively. The tests were carried out for $\varphi$ = 0$^\circ$, 15$^\circ$, 30$^\circ$, 45$^\circ$, and 60$^\circ$.}
		\begin{tabular}{c c c c c c c c}
			\hline\hline
			Case & $m_1$  & $m_2$ & $U_{if}$ & $h_{if}$ &  $\overline{U}/U_{if}$ & $Re_D$ & $Re_d$ \\
			$\ldots$ & $g$ & $g$ & $m/s$ & $mm$ & $\ldots$ & $\ldots$ & $\ldots$ \\
			\hline
			$1$ & $0.3$ & $0$ & $0.63$ & $24$ & $2.5$ & $311$ & $52$ \\
			$2$ & $0.3$ & $0$ & $0.63$ & $24$ & $2.8$ & $350$ & $58$ \\
			$3$ & $0.5$ & $0$ & $0.63$ & $41$ & $2.5$ & $311$ & $52$ \\
			$4$ & $0.5$ & $0$ & $0.63$ & $41$ & $2.8$ & $350$ & $58$ \\
			$5$ & $0.7$ & $0$ & $0.63$ & $56$ & $2.5$ & $311$ & $52$ \\
			$6$ & $0.7$ & $0$ & $0.63$ & $56$ & $2.8$ & $350$ & $58$ \\
			$7$ & $0.2$ & $0.4$ & $0.63$ & $36$ & $2.5$ & $311$ & $52$ \\
			$8$ & $0.2$ & $0.4$ & $0.63$ & $36$ & $2.8$ & $350$ & $58$ \\
			$9$ & $0.2$ & $0.5$ & $0.63$ & $41$ & $2.5$ & $311$ & $52$ \\
			$10$ & $0.2$ & $0.5$ & $0.63$ & $41$ & $2.8$ & $350$ & $58$ \\
			$11$ & $0.2$ & $0.7$ & $0.78$ & $51$ & $2.0$ & $311$ & $52$ \\
			$12$ & $0.2$ & $0.7$ & $0.78$ & $51$ & $2.3$ & $350$ & $58$ \\
			$13$ & $0.3$ & $0.4$ & $0.71$ & $44$ & $2.2$ & $311$ & $52$ \\
			$14$ & $0.3$ & $0.4$ & $0.71$ & $44$ & $2.5$ & $350$ & $58$ \\
			$15$ & $0.3$ & $0.5$ & $0.78$ & $49$ & $2.0$ & $311$ & $52$ \\
			$16$ & $0.3$ & $0.5$ & $0.78$ & $49$ & $2.3$ & $350$ & $58$ \\
			$17$ & $0.3$ & $0.7$ & $0.78$ & $59$ & $2.0$ & $311$ & $52$ \\
			$18$ & $0.3$ & $0.7$ & $0.78$ & $59$ & $2.3$ & $350$ & $58$ \\
			\hline\hline
		\end{tabular}
		\label{tab_exp}
	\end{center}
\end{table}

All experiments were performed at ambient temperature within 21 and 23 $^\circ$C and relative humidity between 46 and 77\%. The velocity of incipient fluidization $U_{if}$ was measured as the minimum velocity required to observe motion of particles, corresponding then to the inception of bed expansion \cite{Zhang}, and is used as an approximation of that for minimum fluidization ($U_{mf}$), such as done by Refs. \cite{Cunez3, Cunez5, Oliveira}. We note that $U_{if}$ (obtained from image processing) can differ from $U_{mf}$ (obtained from pressure drop) because of the high friction and adhesion forces in MFBs, which induce large fluidization-defluidization hysteresis and considerable deviations from correlations for larger (regular) beds \cite{Zhang}. For bidisperse beds, we considered  $U_{if}$ as the onset of fluidization of the top layer (following Formisani et al. \cite{formisani_fluidization_2008}).

A high-speed camera was placed perpendicularly to the tube-pivoting plane to acquire images of the bed. The camera was of complementary metal-oxide-semiconductor (CMOS) type with maximum resolution of 2560 px $\times$ 1600 px at 800 Hz, and was assembled with a $60$-mm-focal-distance lens of F2.8 maximum aperture. The region of interest (ROI) was fixed to 2560 px $\times$ 128 px because of the bed's narrow profile, and the acquiring frequency to 1000 Hz (the reduced ROI allowed an increase in the acquiring frequency). The field of view varied between 213 mm $\times$ 11 mm and 118 mm $\times$ 6 mm, corresponding to resolutions within 12 and 22 px/mm. In order to avoid beating between the camera and lighting, the illumination system consisted of lamps of light-emitting diode (LED) branched to a continuous-current source.

\subsection{Image processing}

Once the test run was recorded, we stored the movie frames as individual images in a computer for being processed later. Image processing began by applying a Gaussian filter for slightly blurring the images in order to reduce noise. Afterward, we applied a threshold filter to distinguish brighter and darker pixels in images, which we associate with grains and background, respectively. The scale of images in physical units (mm) was then determined by correlating the width of the bed (in px) with the tube diameter $D$ (in mm), and the bed height $H$ was measured as the highest longitudinal position (Fig. \ref{fig_layout}a) where a bright pixel could be found in the tube. 

Individual grains were identified by searching for local maxima in the images, and afterward, filtering by size and neighbor distance, the grain radius was determined and false positives eliminated. This approach was necessary because of the relatively low resolution of particles (around $10$ pixels per diameter) and high compactness of the bed, which did not allow the use of more common edge-detection algorithms. The motion of individual grains was determined by minimizing the total distance traveled by all tracked particles between consecutive frames, for which we made use of an Auction algorithm \cite{bertsekas_auction_1992}. In addition, we used a Kalman filter \cite{Kalman} to mitigate (compensate) frames in which a particle is not identified. This tracking approach allowed for the computation of the instantaneous velocities of grains. When the two species were present (bidisperse beds), the identification was done by using the average brightness of each particle, since zirconium appears brighter than glass in the images. We applied a threshold to delimit plugs from bubbles, and eliminated false positives by considering only plugs whose lengths were larger than one grain diameter. The plug length $\lambda$ was then computed as the longitudinal distance between its borders (Fig. \ref{fig_layout}a), and the plug celerity $C$ by tracking its top border along time (i.e., multiplying the value of the longitudinal position by the camera frequency). We generated synthetic images of granular beds (with the same characteristics of those from experiments) in which we knew the exact values of particle velocities (by imposing the displacement of each grain for a given time step between images). By processing such images with our code, we found an average error of 0.6\%, which we consider acceptable.

\section{RESULTS}
\label{sec:results}

\subsection{Monodisperse beds}

As in C\'u\~nez and Franklin \cite{Cunez5}, vertically-aligned monodisperse beds generated alternating high- and low-compactness regions (plugs and bubbles, respectively) occupying the entire tube cross section. Those structures were measured and analyzed in that work, so that we will not report them again here. Instead, we present next the bed behavior when inclined at different angles, and in the following section the behavior of bidisperse beds.

\begin{figure}[h!]
	\centering
	\includegraphics[width=0.99\textwidth]{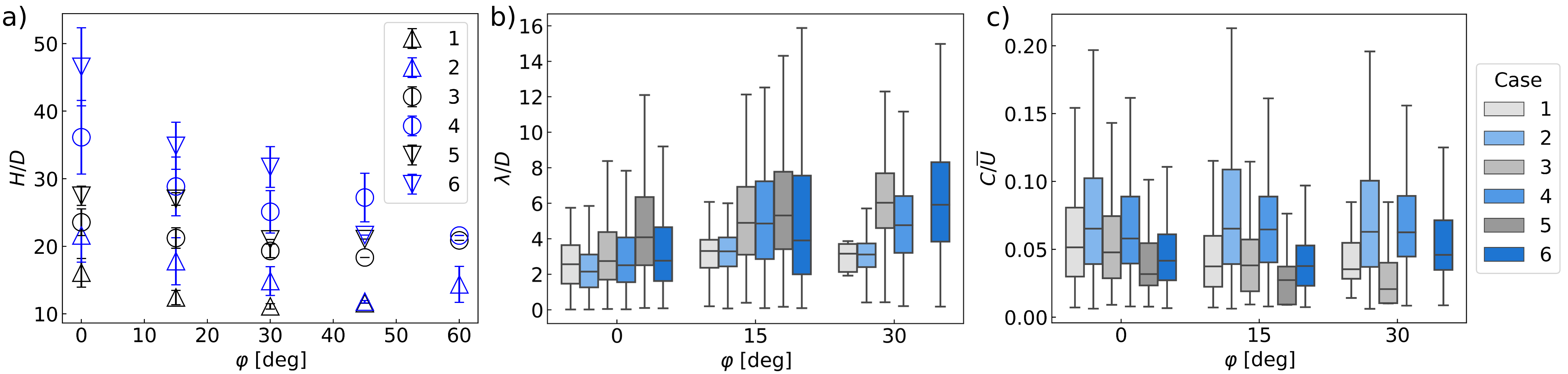}
	\caption{(a) Time-averaged bed height $H$ of monodisperse cases 1 to 6 (glass spheres) as a function of the inclination angle $\varphi$ (with respect to gravity). Symbols are listed in the key (same shape corresponds to a given $m_T$ and same color to a given fluid velocity $\overline{U}$), and the error bars correspond to the standard deviation of measurements. (b) Plug length $\lambda$ and (c) plug celerity $C$ as functions of the inclination angle $\varphi$. Values are normalized by the tube diameter $D$ and cross-sectional mean velocity $\overline{U}$, and graphics are presented as box plots for showing how $\lambda$ and $C$ are distributed. Colors correspond to different cases, as listed in the key.}
	\label{fig:height_mono}
\end{figure}

We begin with the variation of the bed height as a function of the inclination angle $\varphi$ (with respect to gravity), which we present in Fig. \ref{fig:height_mono}a in terms of the time-averaged bed height $H$. Figure \ref{fig:height_mono}a shows that the bed expansion decreases non-monotonically with $\varphi$, until fluidization is stopped at angles close to $\varphi$ = 60$^\circ$. During the decrease in expansion, we observed the formation of a channel in the upper part of the bed (through which the air flowed more freely). We can observe a slight increase in $H$ for some cases when $\varphi$ approaches 60$^\circ$, which occurred due to the settling of grains on the lower part of the tube wall. These observations are, in a certain way, in agreement with those reported by Del Pozo et al. \cite{del_pozo_effect_1992} (for the small angles tested in their study). We note that the experimental points in Fig. \ref{fig:height_mono} present higher errors (standard deviations) when plugs and bubbles are present in the bed, producing then relatively high oscillations around a mean value (and are not due to measurement uncertainties).

Figures \ref{fig:height_mono}b and \ref{fig:height_mono}c show the plug length $\lambda$ and celerity $C$ as functions of the inclination angle $\varphi$, and they are normalized by the tube diameter $D$ and cross-sectional mean velocity $\overline{U}$, respectively. In the graphics, the box plots show how $\lambda$ and $C$ are distributed, and we observe that average values of $\lambda$ and $C$ remain roughly constant with $\varphi$, while their distributions spread around the mean values. The increase in spreading basically accounts for plugs becoming surface waves as the bed deviates from the vertical position, being strongly related to the decrease in $H$ and the corresponding errors as $\varphi$ increases. The graphics do not show values for $\varphi$ greater than 30$^\circ$ because plugs no longer exist at those inclinations. In addition, case 5 of Tab. \ref{tab_exp} did not generate plugs at 30$^\circ$, and for this reason the respective $\lambda$ and $C$ are not shown for this inclination in Figs. \ref{fig:height_mono}b and \ref{fig:height_mono}c.

\subsection{Bidisperse beds}

\begin{figure}[h!]
	\centering
	\includegraphics[width=0.99\textwidth]{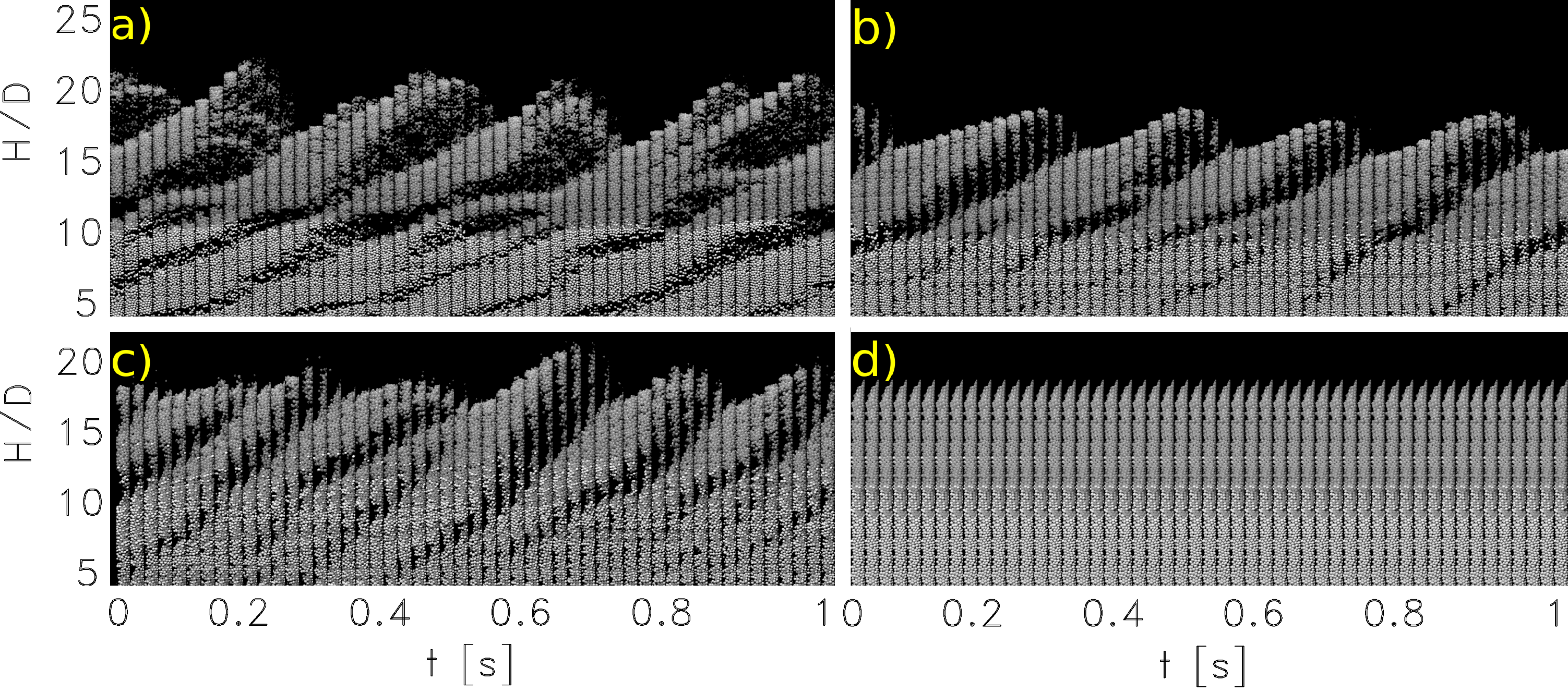}\\
	\caption{Snapshots placed side by side of a bidisperse fluidized bed with $m_1$ = 0.2 g, $m_2$ = 0.5 g, $\overline{U}/U_{if}$ = 2.8 (case 10 of Tab. \ref{tab_exp}), and $\varphi$ = (a) 0$^\circ$, (b)15$^\circ$, (c) 30$^\circ$ and (d) 45$^\circ$, with a time interval of 20 ms between frames and 1 s of total time. The images have a reasonable resolution, so that the grains can be individually distinguished by zooming in. Panels a--d are available in larger size in the supplementary material. Multimedia available online.}
	\label{fig:bidisperse_snapshot}
\end{figure}

\begin{figure}[h!]
	\centering
	\includegraphics[width=0.6\textwidth]{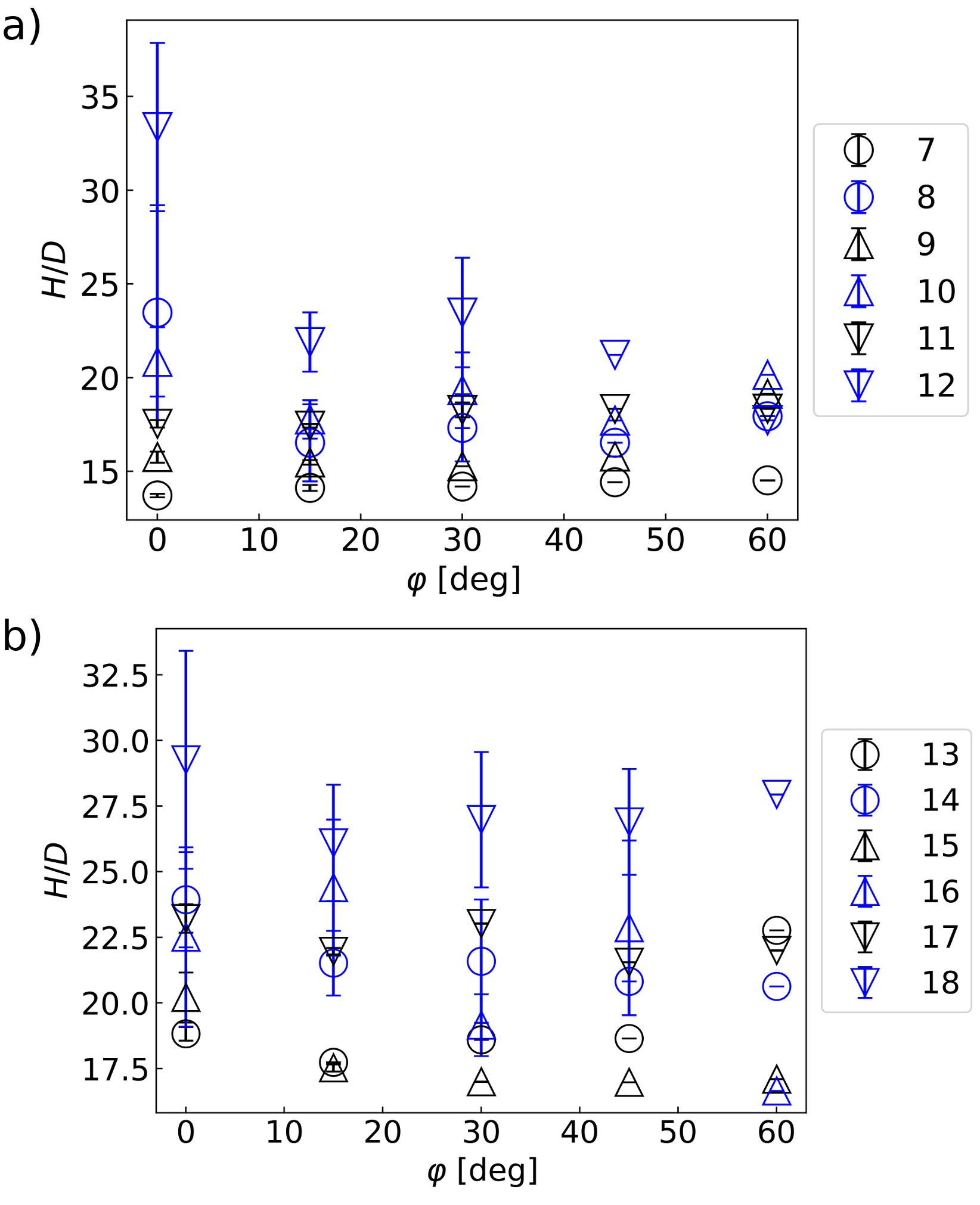}
	\caption{Time-averaged bed height $H$ of bidisperse cases (a) 7 to 12 and (b) 13 to 18 as a function of the inclination angle $\varphi$. Symbols are listed in the key (same shape corresponds to a given $m_T$ and same color to a given fluid velocity $\overline{U}$), and error bars correspond to the standard deviation of measurements.}
	\label{fig:height_bi}
\end{figure}

Figure \ref{fig:bidisperse_snapshot} (Multimedia view) shows snapshots placed side by side of a bidisperse fluidized bed at different inclinations (panels a--d for $\varphi$ = 0$^\circ$, 15$^\circ$, 30$^\circ$ and 45$^\circ$, respectively). For this bed composition  ($m_1$ = 0.2 g and $m_2$ = 0.5 g), we observe a decrease in the bed height as $\varphi$ increases, with a decrease in the size of bubbles (but not necessarily in the plug size) and a corresponding decrease in oscillations of $H$. As $\varphi$ reaches 45$^\circ$, great part of grains settle over the lower portion of the tube wall (since from this angle on, the bed approaches the horizontal position), and plugs and bubbles give place to surface waves that propagate on the part of the bed that is farther from the tube wall (bed surface). For this reason, error bars are lower for higher angles in Fig. \ref{fig:height_bi}.

At the grain scale, Fig. \ref{fig:bidisperse_snapshot} shows that the segregation pattern changes with inclination, with the appearance of a mixing layer that increases until $\varphi$ reaches approximately 45$^\circ$ (as we will see next), from which angle fluidization decreases and, consequently, mixing decreases. The existence of an optimal angle for mixing in bidisperse beds is an interesting behavior that can be explored in industrial processes, and of which we investigate the details in the following.

\begin{figure}[h!]
	\centering
	\includegraphics[width=0.99\textwidth]{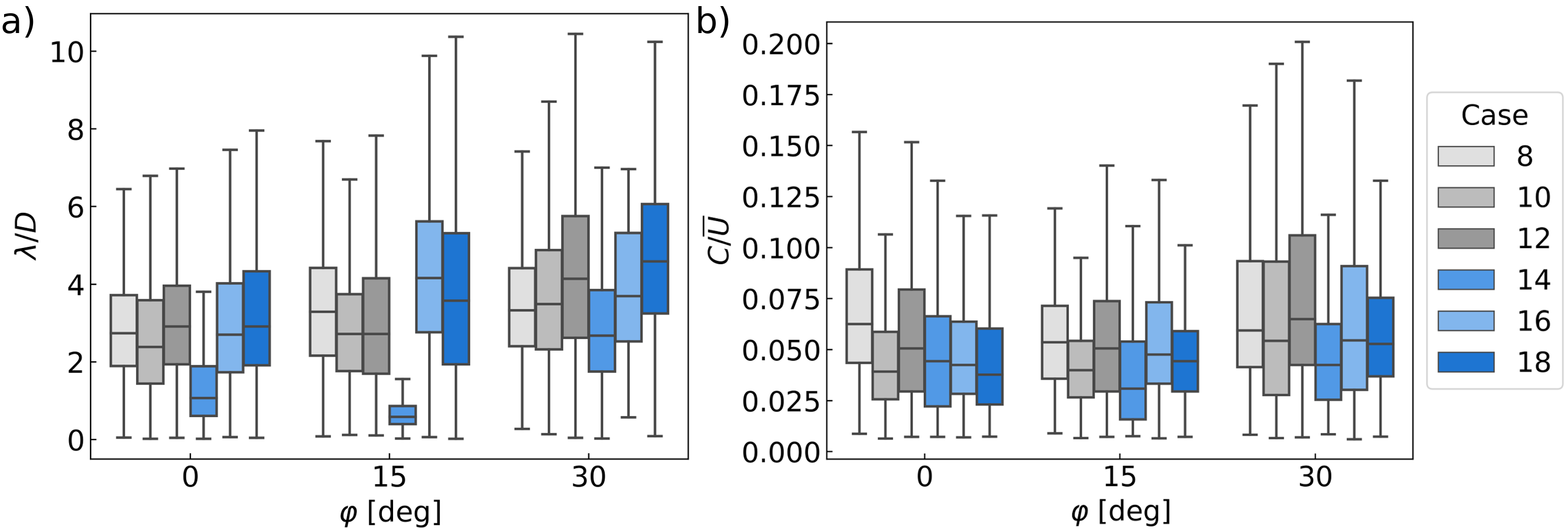}
	\caption{(a) Plug length $\lambda$ and (b) plug celerity $C$ as functions of the inclination angle $\varphi$. Values are normalized by the tube diameter $D$ and cross-sectional mean velocity $\overline{U}$, and graphics are presented as box plots for showing explicitly how $\lambda$ and $C$ are distributed. Colors correspond to different cases, as listed in the key.}
	\label{fig:plugs_bi}
\end{figure}

We begin by investigating the bed behavior at the bed scale. Figure \ref{fig:height_bi} shows the time-averaged bed height $H$ as a function of the inclination angle $\varphi$ for the bidisperse beds tested. We observe that the decrease in bed expansion as $\varphi$ increases is much less pronounced in most cases (in particular when more zirconium particles are present) than in monodisperse beds. However, the general behavior is similar: bed expansion decreases until $\varphi$ reaches an angle that depends on the bed composition, and, from this point on, $H$ increases. Fluidization is finally stopped at angles close to 60$^\circ$. As for Fig. \ref{fig:height_mono}, we note that experimental points present higher errors (standard deviations) when plugs and bubbles are present in the bed. Figures \ref{fig:plugs_bi}a and \ref{fig:plugs_bi}b show the plug length $\lambda$ and celerity $C$ as functions of the inclination angle $\varphi$, and are presented as box plots. As for the modisperse case, the average values of $\lambda$ and $C$ remain roughly constant with $\varphi$, while the distributions spread (the latter accounting for plugs becoming surface waves). The graphics do not show values for $\varphi$ greater than 30$^\circ$ because plugs no longer exist at those inclinations.

\begin{figure}[h!]
	\begin{center}
		\includegraphics[width=0.6\linewidth]{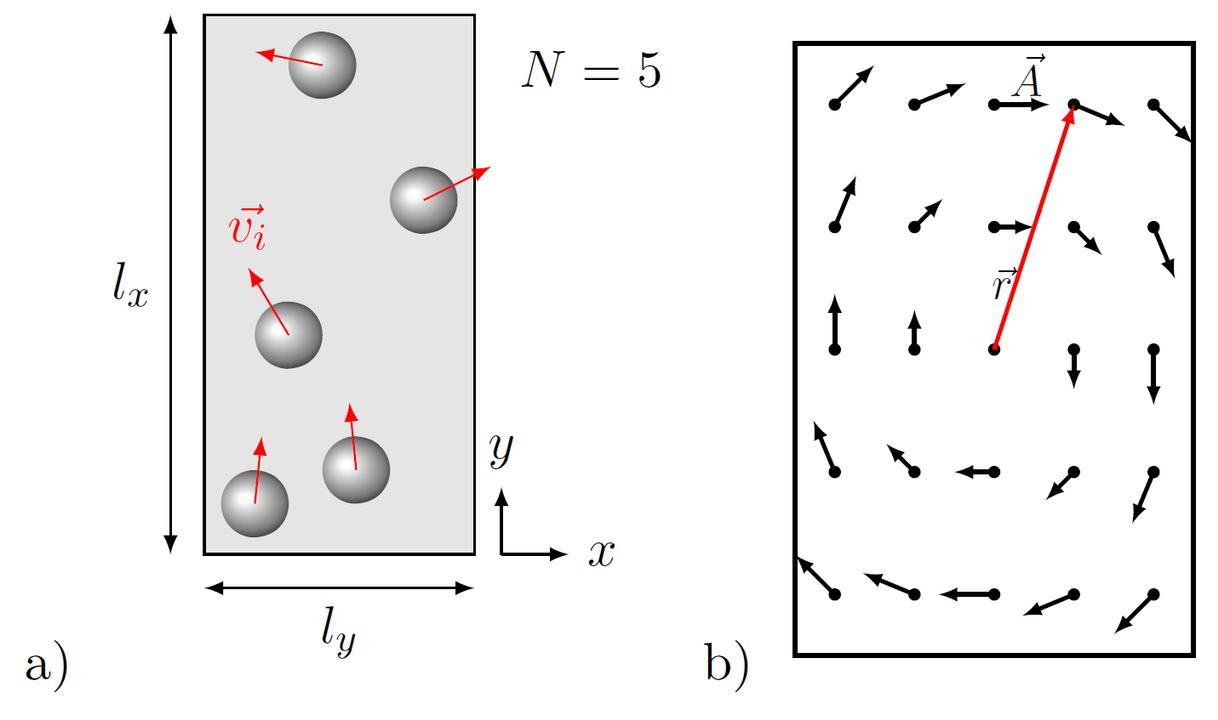}
	\end{center}
	\caption{Diagrams showing (a) regions in which granular temperature $\theta$ was computed based on ensemble averages and (b) definition of the bed moment $\vec{L}$.}
	\label{fig:diagram}
\end{figure}

\begin{figure}[h!]
	\centering
	\includegraphics[width=0.6\textwidth]{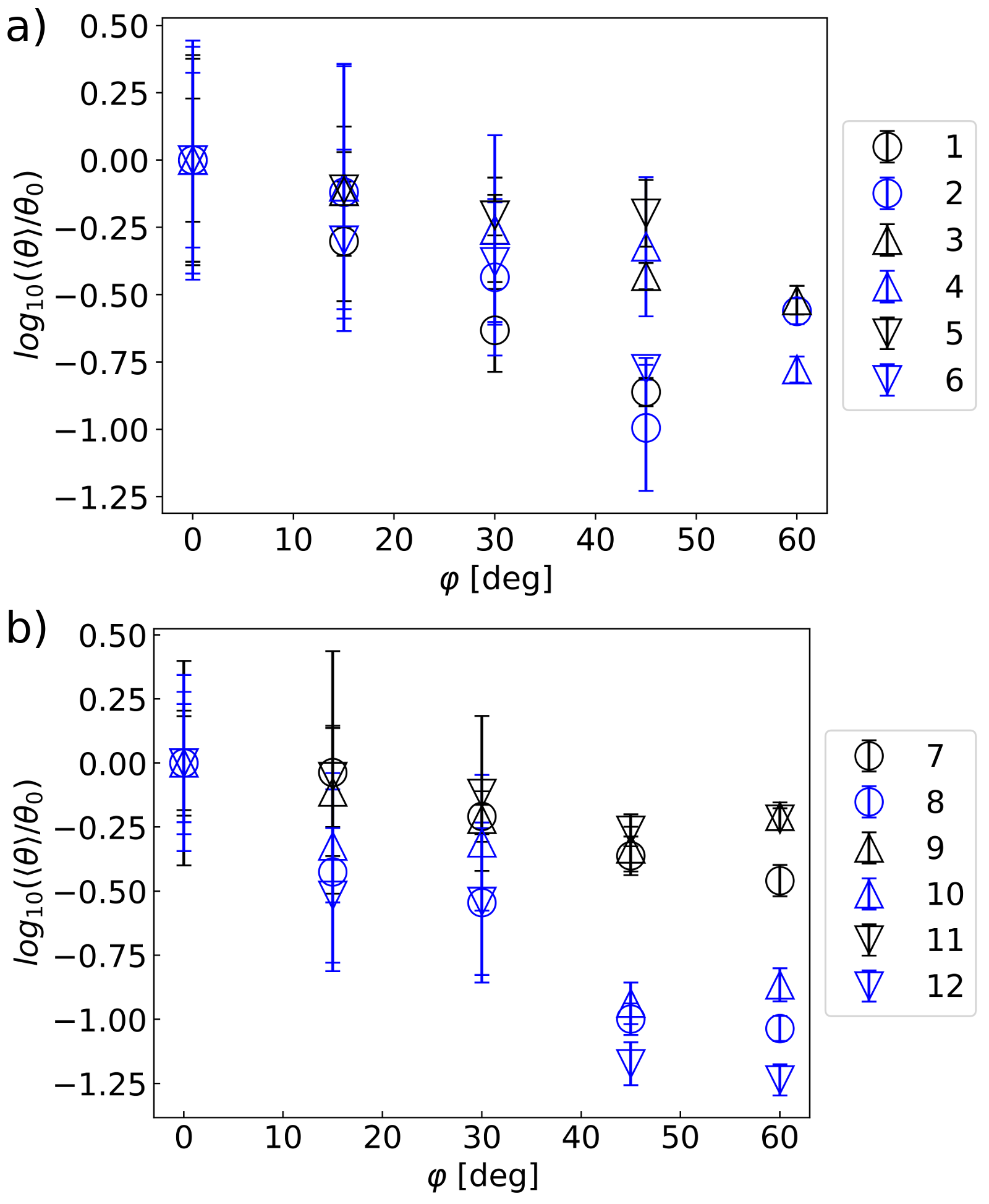}
	\caption{Space-time averages of the granular temperature for the entire bed $\left< \theta \right>$ as a function of the inclination angle $\varphi$. The ordinates present the logarithm of $\left< \theta \right>$ (normalized by the mean value for the vertical bed $\theta_0$) in order to accentuate differences. Symbols are listed in the key (same shape corresponds to a given $m_T$ and same color to a given fluid velocity $\overline{U}$), and error bars correspond to the standard deviation of measurements.}
	\label{fig:gran_temp}
\end{figure}

We now investigate the bed behavior at the grain scale by computing the granular temperature, circulation of particles, and degree of mixing. We note that we only had optical access to grains in contact with the tube walls (the core of granular plugs, where local packing fraction is expected to be higher, was not visible). Therefore, those quantities were computed in two dimensions, for which we used the projection of the cylindrical surface on a Cartesian plane. For the granular temperature, we first divided the bed into regions measuring 1.2$d$ in the horizontal and 10$d$ in the vertical direcion, and computed the ensemble average $\left< v \right>$ of the grains' velocity for each cell,

\begin{equation}
	\left< v_k \right> = \frac{1}{N} \sum_{i=1}^{N} v_{ki} \text{ , with } k = x,y \nonumber \,\,,
	\label{eq:avg_velocity}
\end{equation}

\noindent where the subscripts $k$ and $i$ stand, respectively, for the component ($x$ or $y$) and considered particle, and $N$ is the total number of grains. Afterward, the local granular temperature $\theta$ was computed by an ensemble average,

\begin{equation}
	\theta = \frac{1}{2} \sum_{k=x,y} \frac{1}{N_{cell}} \sum_{i=1}^{N_{cell}} [v_{ki}^2 - \left< v_k \right> ^2] \,\,,
	\label{eq:granular_temp}
\end{equation}

\noindent where $N_{cell}$ is the number of particles in the considered cell (Fig.\ref{fig:diagram}a shows an illustrating diagram). Finally, we computed the space-time average of the granular temperature for the entire bed $\left< \theta \right>$ and duration of tests, for each case tested. This average indicates the degree of agitation at the level of grains within the entire bed, and they are shown in Fig. \ref{fig:gran_temp} in dimensionless form (normalized by the mean value for the vertical bed $\theta_0$). A figure in dimensional form and space-time diagrams of the granular temperature $\theta$ for different angles are available in the supplementary material. This figure shows that the average granular temperature $\left< \theta \right>$ decreases with increasing $\varphi$, which corresponds to lower levels of fluidization as the bed is inclined toward a horizontal flow. For increasing $\varphi$, an increasing number of grains settle at the lower part of the tube wall, with the formation of an air channel in the region closer to the upper part of the tube wall \cite{odea_effect_1990}.

Since the agitation of grains decreases with $\varphi$, the motion of grains must present a circulatory component for an increase in mixing at intermediate angles (as observed in Fig. \ref{fig:bidisperse_snapshot}). This can be investigated by computing the volumetric flux of particles \cite{jiang_color-ptv_2018},  calculated in pre-defined cells throughout the bed by using the velocity of each particle, accounting then for the flux of solids in terms of volume. Because in our images only grains in contact with the tube wall are visible, we compute a flux in terms of area: In each grid cell, the area flux $\vec{A}$ is computed as the sum of the particle's area ($\pi r_i^2$) multiplied by its velocity, and divided by the cell's length ($l_k$) perpendicular to the velocity component $k$, as shown in Eq. \ref{eq:area_flux},

\begin{equation}
	A_k = \frac{1}{l_k} \sum_{i=1}^{N} [v_{ki} ~\pi r_i^2] \, \text{ , with } k = x,y \,\,.
		\label{eq:area_flux}
\end{equation}

\noindent A parameter that further represents the circulation of solids is the bed moment $\vec{L}$, computed as the cross product of the distance measured from the center of rotation $\vec{r}$ by the area flux $\vec{A}$, summed over all $j$ regions in the bed,

\begin{equation}
	L = \sum_{i=1}^{j} [\vec{r} \times \vec{A}] \cdot \vec{e}_{z} \,\,,
	\label{eq:moment}
\end{equation}

\noindent where $\vec{e}_{z}$ is the unity vector perpendicular to the considered plane and $L$ = $\vec{L}\cdot\vec{e}_{z}$ (Fig.\ref{fig:diagram}b shows an illustrating diagram).

\begin{figure}[h!]
	\centering
	\includegraphics[width=0.6\textwidth]{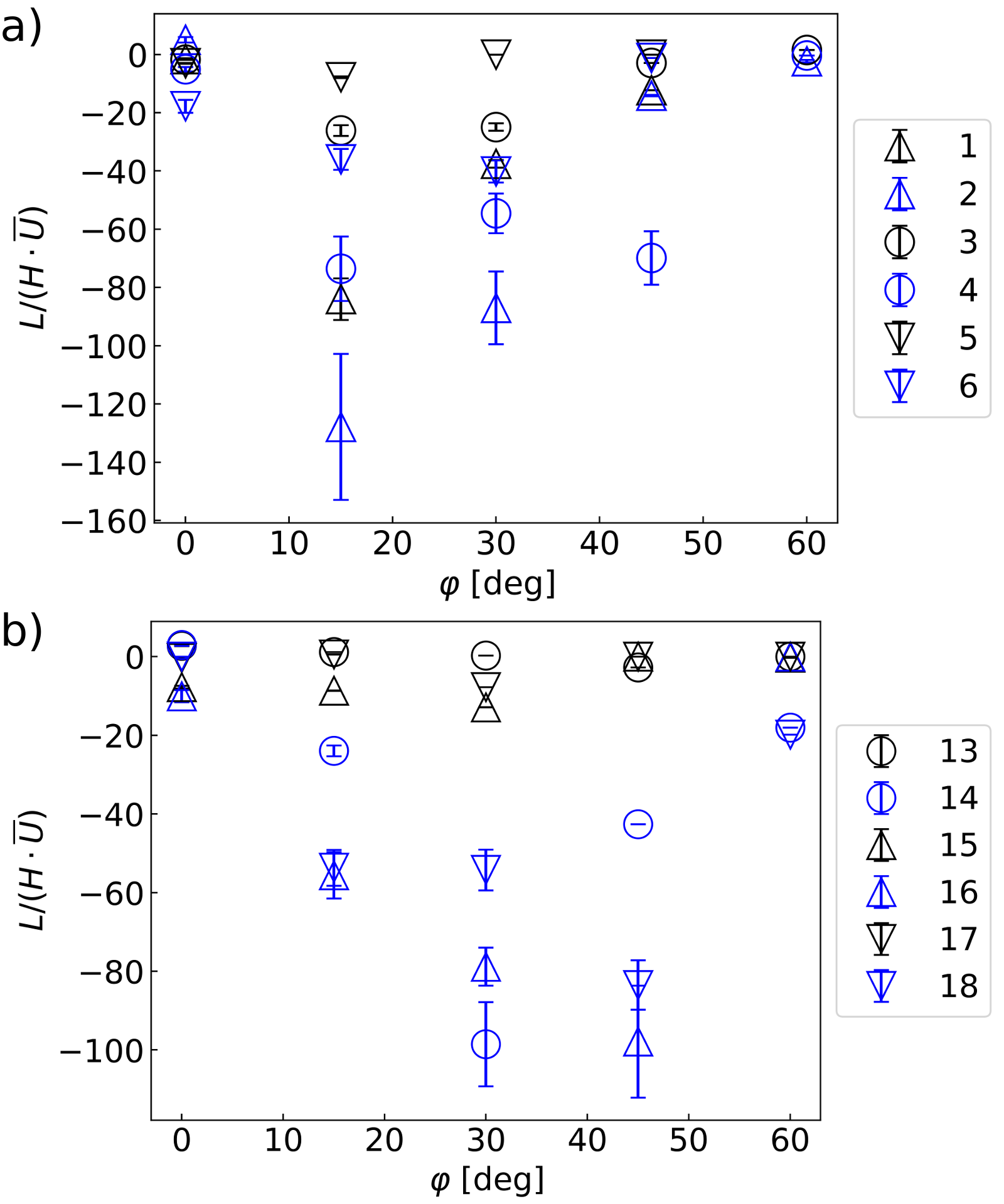}
	\caption{Magnitude of the bed moment $L$ normalized by the bed height $H$ and cross-sectional velocity of air $\overline{U}$ as a function of the inclination angle $\varphi$, for (a) monodisperse beds (b) bidisperse beds with $m_1$ = 0.3 g. Symbols are listed in the key (same shape corresponds to a given $m_T$ and same color to a given fluid velocity $\overline{U}$), and error bars correspond to the standard deviation of measurements.}
	\label{fig:bed_moment}
\end{figure}

Figure \ref{fig:bed_moment} presents the magnitude of the bed moment $L$ normalized by the bed height $H$ and cross-sectional velocity of air $\overline{U}$ as a function of the inclination angle $\varphi$, for both monodisperse and bidisperse beds. The values are negative because of axes orientation, since the bed inclinations promoted clockwise rotations (counterclockwise rotations would imply positive values). Therefore, absolute values of $L$ in Fig. \ref{fig:bed_moment} correspond to the momentum strength. Although values vary with the bed composition and size, we observe that stronger circulations of grains in monodisperse beds occur for angles around 10$^\circ$--20$^\circ$, while in bidisperse beds they occur for angles within 30$^\circ$--50$^\circ$, i.e., closer to the horizontal position when compared to the monodisperse case. We expect, thus, stronger mixing in the bidisperse cases when inclinations are around 30$^\circ$--50$^\circ$ with respect to gravity, which is corroborated by direct measurements of mixing (presented next). In terms of strength, the moduli of $L/(H \overline{U})$ reach higher values in monodisperse beds ($\approx$ 140) in comparison with bidisperse beds ($\approx$ 100). Graphics of the area flux along the bed, for different inclinations, are available in the supplementary material \cite{Supplemental3}, showing in detail the circulation patterns (vortices) of groups of particles. One of the graphics shows that patterns are not uniform along the bed, but consist in vortices that circulate grains between adjacent plugs and propagate in the main flow direction.

\begin{figure}[h!]
	\centering
	\includegraphics[width=0.9\textwidth]{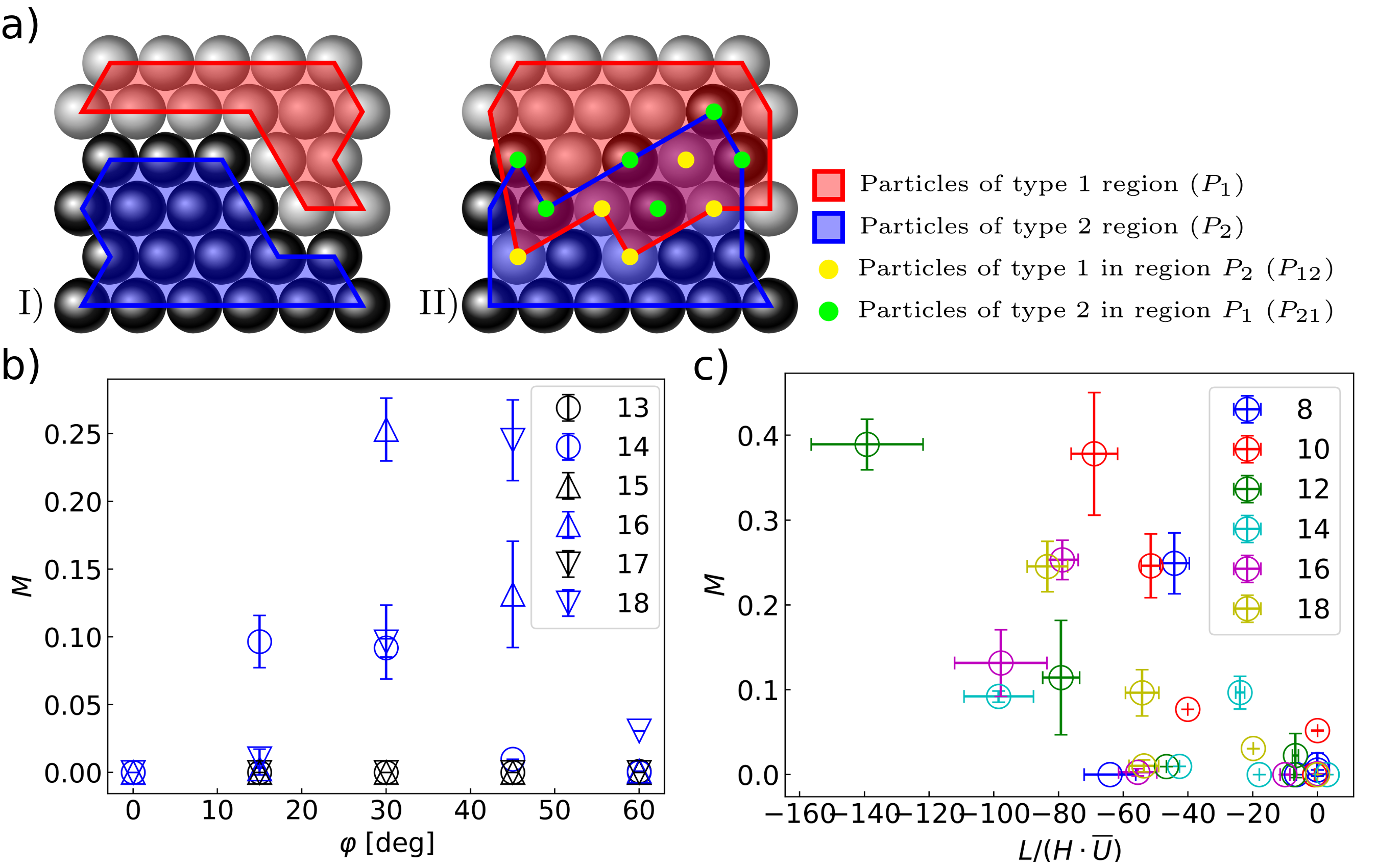}
	\caption{(a) Diagram indicating how mixing is considered for computations. (b) Mixing ratio $M$ as a function of the inclination angle $\varphi$, for bidisperse beds with $m_1$ = 0.3 g. Symbols are listed in the key (same shape corresponds to a given $m_T$ and same color to a given fluid velocity $\overline{U}$), and error bars correspond to the standard deviation of measurements. (c) Mixing ratio $M$ as a function of the dimensionless bed moment $L$. Multimedia available online.}
	\label{fig:mixing_inclination}
\end{figure}

To estimate the degree of mixing of different species, a region is traced to surround every particle of the corresponding type and a frontier is established. Then, only particles of a given species trespassing the frontier of the other one are counted for computing the degree of mixing, as shown in Fig. \ref{fig:mixing_inclination}a. Then, the mixing ratio $M$ is computed as the ratio between the number of particles in the mixing layer and the total number of particles,

\begin{equation}
	M = \frac{P_{21} + P_{12}}{P_1 + P_2} \,\,,
	\label{eq:mixing}
\end{equation}

\noindent where $P_m$ is the number of particles of species $m$ and $P_{mn}$ is the number of particles of species $m$ trespassing the frontier of species $n$. Figure \ref{fig:mixing_inclination}b (Multimedia view) shows the mixing ratio $M$ as a function of the inclination angle $\varphi$ for some bidisperse beds. We observe that, indeed, stronger mixing occurs for angles within 30$^\circ$--50$^\circ$, where the mixing ratio $M$ reaches values up to 0.25. Finally, in order to verify the relation between circulation of grains and mixing, we plotted the mixing ratio $M$ as a function of the bed moment $L$, which we show in Fig. \ref{fig:mixing_inclination}c in dimensionless form. Although with some dispersion, we observe a tendency of higher mixing for higher bed moment (in modulus).

Further investigations can carry out numerical computations, in particular Euler-Lagrange simulations (computational fluid dynamics - discrete element method, for example), for measuring in detail the motion of all grains within the bed. In addition, the results would be in three dimensions, so that grain-scale data such as the granular temperature would be more accurate. Because of the small number of particles in the bed, we believe that the Euler-Euler approach is not suited for this problem.

In summary, for the range of parameters investigated in this study, mixing in bidispese MFBs is enhanced for inclination angles within 30$^\circ$--50$^\circ$ with respect to gravity. The non-monotonic behavior results from a competition between worse fluidization (lower agitation) and stronger circulation as inclination deviates from the vertical orientation. The present results can be useful for improving mechanical, chemical and pharmaceutical processes in which mixing in MFBs is necessary.

\section{CONCLUSIONS}
\label{sec:conclusions}

In this paper, we investigated experimentally the dynamics of mono and bidisperse gas-solid MFBs under different inclinations. In the experiments, MFBs of the very-narrow type ($D/d$ = 6, i.e., highly confined) were filmed with a high-speed camera and the images processed for obtaining measurements down to the grain scale. For monodisperse beds, we found that plugs become surface waves as the bed deviates from the vertical position, with plugs no longer existing for deviation angles $\varphi$ higher than 30$^\circ$. Within 0$^\circ$ $\leq$ $\varphi$ $\lessapprox$ 30$^\circ$, average values of plug length $\lambda$ and celerity $C$ remain roughly constant. We observed the same behavior for bidispese MFBs and, in addition, found that: (i) the degree of segregation is larger for vertical beds; (ii) mixing varies non-monotonically with the bed inclination, with an optimal angle within 30$^{\circ}$ $\leq$ $\varphi$ $\leq$ 50$^{\circ}$ where the mixing ratio $M$ reaches values up to 0.25; (iii) granular temperature and fluidization decrease with the increase of $\varphi$; (iv) circulation of grains varies non-monotonically with the increase of $\varphi$, with the magnitude of the normalized bed moment $L/(H \overline{U})$ (associated with circulation) reaching values up to 100 in the bidisperse case when 30$^{\circ}$ $\leq$ $\varphi$ $\leq$ 50$^{\circ}$; (v) as a consequence, the mixing layer results from the competition between segregation by kinetic sieving and circulation promoted by the fluid flow. Finally, we presented a diagram showing how mixing $M$  varies with the bed moment $L$, evincing a direct relation between both, and a diagram of mixing $M$ vs. deviation angle $\varphi$ that can be used for determining optimal mixing. Our results can be useful to improve mechanical, chemical and pharmaceutical processes in which mixing in MFBs is necessary, such as capturing $CO_2$, treating wastwater, and producing tablets, pills and vaccines in powder form, for example.

\section*{AUTHOR DECLARATIONS}
\noindent \textbf{Conflict of Interest}

The authors have no conflicts to disclose

\section*{SUPPLEMENTARY MATERIAL}
See the supplementary material for additional figures and supporting information.

\section*{DATA AVAILABILITY}

The data that support the findings of this study are openly available in Mendeley Data at\\ \url{https://doi.org/10.17632/v9c5bsfz7t.1}\cite{Supplemental3}.

\begin{acknowledgments}
The authors are grateful to the S\~ao Paulo Research Foundation -- FAPESP (Grant Nos. 2018/14981-7 and 2022/08321-0) and to the Conselho Nacional de Desenvolvimento Cient\'ifico e Tecnol\'ogico -- CNPq (Grant No. 405512/2022-8) for the financial support provided.
\end{acknowledgments}

\bibliography{references}

%aipnum4-2.bst 2019-01-14 (MD) hand-edited version of apsrev4-1.bst
\begin{thebibliography}{54}%
\makeatletter
\providecommand \@ifxundefined [1]{%
 \@ifx{#1\undefined}
}%
\providecommand \@ifnum [1]{%
 \ifnum #1\expandafter \@firstoftwo
 \else \expandafter \@secondoftwo
 \fi
}%
\providecommand \@ifx [1]{%
 \ifx #1\expandafter \@firstoftwo
 \else \expandafter \@secondoftwo
 \fi
}%
\providecommand \natexlab [1]{#1}%
\providecommand \enquote  [1]{``#1''}%
\providecommand \bibnamefont  [1]{#1}%
\providecommand \bibfnamefont [1]{#1}%
\providecommand \citenamefont [1]{#1}%
\providecommand \href@noop [0]{\@secondoftwo}%
\providecommand \href [0]{\begingroup \@sanitize@url \@href}%
\providecommand \@href[1]{\@@startlink{#1}\@@href}%
\providecommand \@@href[1]{\endgroup#1\@@endlink}%
\providecommand \@sanitize@url [0]{\catcode `\\12\catcode `\$12\catcode
  `\&12\catcode `\#12\catcode `\^12\catcode `\_12\catcode `\%12\relax}%
\providecommand \@@startlink[1]{}%
\providecommand \@@endlink[0]{}%
\providecommand \url  [0]{\begingroup\@sanitize@url \@url }%
\providecommand \@url [1]{\endgroup\@href {#1}{\urlprefix }}%
\providecommand \urlprefix  [0]{URL }%
\providecommand \Eprint [0]{\href }%
\providecommand \doibase [0]{https://doi.org/}%
\providecommand \selectlanguage [0]{\@gobble}%
\providecommand \bibinfo  [0]{\@secondoftwo}%
\providecommand \bibfield  [0]{\@secondoftwo}%
\providecommand \translation [1]{[#1]}%
\providecommand \BibitemOpen [0]{}%
\providecommand \bibitemStop [0]{}%
\providecommand \bibitemNoStop [0]{.\EOS\space}%
\providecommand \EOS [0]{\spacefactor3000\relax}%
\providecommand \BibitemShut  [1]{\csname bibitem#1\endcsname}%
\let\auto@bib@innerbib\@empty
%</preamble>
\bibitem [{\citenamefont {C\'u\~nez}\ and\ \citenamefont
  {Franklin}(2019)}]{Cunez}%
  \BibitemOpen
  \bibfield  {author} {\bibinfo {author} {\bibfnamefont {F.~D.}\ \bibnamefont
  {C\'u\~nez}}\ and\ \bibinfo {author} {\bibfnamefont {E.}~\bibnamefont
  {Franklin}},\ }\bibfield  {title} {\enquote {\bibinfo {title} {Plug regime in
  water fluidized beds in very narrow tubes},}\ }\href@noop {} {\bibfield
  {journal} {\bibinfo  {journal} {Powder Technol.}\ }\textbf {\bibinfo {volume}
  {345}},\ \bibinfo {pages} {234--246} (\bibinfo {year} {2019})}\BibitemShut
  {NoStop}%
\bibitem [{\citenamefont {C\'u\~nez}\ and\ \citenamefont
  {Franklin}(2020{\natexlab{a}})}]{Cunez2}%
  \BibitemOpen
  \bibfield  {author} {\bibinfo {author} {\bibfnamefont {F.~D.}\ \bibnamefont
  {C\'u\~nez}}\ and\ \bibinfo {author} {\bibfnamefont {E.~M.}\ \bibnamefont
  {Franklin}},\ }\bibfield  {title} {\enquote {\bibinfo {title} {Mimicking
  layer inversion in solid-liquid fluidized beds in narrow tubes},}\
  }\href@noop {} {\bibfield  {journal} {\bibinfo  {journal} {Powder Technol.}\
  }\textbf {\bibinfo {volume} {364}},\ \bibinfo {pages} {994--1008} (\bibinfo
  {year} {2020}{\natexlab{a}})}\BibitemShut {NoStop}%
\bibitem [{\citenamefont {C\'u\~nez}\ and\ \citenamefont
  {Franklin}(2020{\natexlab{b}})}]{Cunez3}%
  \BibitemOpen
  \bibfield  {author} {\bibinfo {author} {\bibfnamefont {F.~D.}\ \bibnamefont
  {C\'u\~nez}}\ and\ \bibinfo {author} {\bibfnamefont {E.~M.}\ \bibnamefont
  {Franklin}},\ }\bibfield  {title} {\enquote {\bibinfo {title}
  {Crystallization and jamming in narrow fluidized beds},}\ }\href@noop {}
  {\bibfield  {journal} {\bibinfo  {journal} {Phys. Fluids}\ }\textbf {\bibinfo
  {volume} {32}},\ \bibinfo {pages} {083303} (\bibinfo {year}
  {2020}{\natexlab{b}})}\BibitemShut {NoStop}%
\bibitem [{\citenamefont {C\'u\~nez}, \citenamefont {Lima},\ and\ \citenamefont
  {Franklin}(2021)}]{Cunez4}%
  \BibitemOpen
  \bibfield  {author} {\bibinfo {author} {\bibfnamefont {F.~D.}\ \bibnamefont
  {C\'u\~nez}}, \bibinfo {author} {\bibfnamefont {N.~C.}\ \bibnamefont
  {Lima}},\ and\ \bibinfo {author} {\bibfnamefont {E.~M.}\ \bibnamefont
  {Franklin}},\ }\bibfield  {title} {\enquote {\bibinfo {title} {Motion and
  clustering of bonded particles in narrow solid–liquid fluidized beds},}\
  }\href@noop {} {\bibfield  {journal} {\bibinfo  {journal} {Phys. Fluids}\
  }\textbf {\bibinfo {volume} {33}},\ \bibinfo {pages} {023303} (\bibinfo
  {year} {2021})}\BibitemShut {NoStop}%
\bibitem [{\citenamefont {Duru}\ and\ \citenamefont {Guazzelli}(2002)}]{Duru}%
  \BibitemOpen
  \bibfield  {author} {\bibinfo {author} {\bibfnamefont {P.}~\bibnamefont
  {Duru}}\ and\ \bibinfo {author} {\bibfnamefont {{\'E}.}~\bibnamefont
  {Guazzelli}},\ }\bibfield  {title} {\enquote {\bibinfo {title} {Experimental
  investigation on the secondary instability of liquid-fluidized beds and the
  formation of bubbles},}\ }\href@noop {} {\bibfield  {journal} {\bibinfo
  {journal} {J. Fluid Mech.}\ }\textbf {\bibinfo {volume} {470}},\ \bibinfo
  {pages} {359–382} (\bibinfo {year} {2002})}\BibitemShut {NoStop}%
\bibitem [{\citenamefont {Duru}\ \emph {et~al.}(2002)\citenamefont {Duru},
  \citenamefont {Nicolas}, \citenamefont {Hinch},\ and\ \citenamefont
  {Guazzelli}}]{Duru2}%
  \BibitemOpen
  \bibfield  {author} {\bibinfo {author} {\bibfnamefont {P.}~\bibnamefont
  {Duru}}, \bibinfo {author} {\bibfnamefont {M.}~\bibnamefont {Nicolas}},
  \bibinfo {author} {\bibfnamefont {J.}~\bibnamefont {Hinch}},\ and\ \bibinfo
  {author} {\bibfnamefont {{\'E}.}~\bibnamefont {Guazzelli}},\ }\bibfield
  {title} {\enquote {\bibinfo {title} {Constitutive laws in liquid-fluidized
  beds},}\ }\href@noop {} {\bibfield  {journal} {\bibinfo  {journal} {J. Fluid
  Mech.}\ }\textbf {\bibinfo {volume} {452}},\ \bibinfo {pages} {371–404}
  (\bibinfo {year} {2002})}\BibitemShut {NoStop}%
\bibitem [{\citenamefont {Oliveira}, \citenamefont {Borges},\ and\
  \citenamefont {Franklin}(2023)}]{Oliveira}%
  \BibitemOpen
  \bibfield  {author} {\bibinfo {author} {\bibfnamefont {V.~P.~S.}\
  \bibnamefont {Oliveira}}, \bibinfo {author} {\bibfnamefont {D.~S.}\
  \bibnamefont {Borges}},\ and\ \bibinfo {author} {\bibfnamefont {E.~M.}\
  \bibnamefont {Franklin}},\ }\bibfield  {title} {\enquote {\bibinfo {title}
  {Crystallization and refluidization in very-narrow fluidized beds},}\ }\href
  {https://doi.org/10.1063/5.0163555} {\bibfield  {journal} {\bibinfo
  {journal} {Physics of Fluids}\ }\textbf {\bibinfo {volume} {35}},\ \bibinfo
  {pages} {093306} (\bibinfo {year} {2023})}\BibitemShut {NoStop}%
\bibitem [{\citenamefont {Goldman}\ and\ \citenamefont
  {Swinney}(2006)}]{Goldman}%
  \BibitemOpen
  \bibfield  {author} {\bibinfo {author} {\bibfnamefont {D.~I.}\ \bibnamefont
  {Goldman}}\ and\ \bibinfo {author} {\bibfnamefont {H.~L.}\ \bibnamefont
  {Swinney}},\ }\bibfield  {title} {\enquote {\bibinfo {title} {Signatures of
  glass formation in a fluidized bed of hard spheres},}\ }\href@noop {}
  {\bibfield  {journal} {\bibinfo  {journal} {Phys. Rev. Lett.}\ }\textbf
  {\bibinfo {volume} {96}},\ \bibinfo {pages} {145702} (\bibinfo {year}
  {2006})}\BibitemShut {NoStop}%
\bibitem [{\citenamefont {Zhang}\ \emph {et~al.}(2021)\citenamefont {Zhang},
  \citenamefont {Goh}, \citenamefont {Ng}, \citenamefont {Chow}, \citenamefont
  {Wang},\ and\ \citenamefont {Zivkovic}}]{Zhang}%
  \BibitemOpen
  \bibfield  {author} {\bibinfo {author} {\bibfnamefont {Y.}~\bibnamefont
  {Zhang}}, \bibinfo {author} {\bibfnamefont {K.-L.}\ \bibnamefont {Goh}},
  \bibinfo {author} {\bibfnamefont {Y.~L.}\ \bibnamefont {Ng}}, \bibinfo
  {author} {\bibfnamefont {Y.}~\bibnamefont {Chow}}, \bibinfo {author}
  {\bibfnamefont {S.}~\bibnamefont {Wang}},\ and\ \bibinfo {author}
  {\bibfnamefont {V.}~\bibnamefont {Zivkovic}},\ }\bibfield  {title} {\enquote
  {\bibinfo {title} {Process intensification in micro-fluidized bed systems:
  {A} review},}\ }\href@noop {} {\bibfield  {journal} {\bibinfo  {journal}
  {Chem. Eng. Processing - Process Intensif.}\ }\textbf {\bibinfo {volume}
  {164}},\ \bibinfo {pages} {108397} (\bibinfo {year} {2021})}\BibitemShut
  {NoStop}%
\bibitem [{\citenamefont {Qie}\ \emph {et~al.}(2022)\citenamefont {Qie},
  \citenamefont {Alhassawi}, \citenamefont {Sun}, \citenamefont {Gao},
  \citenamefont {Zhao},\ and\ \citenamefont {Fan}}]{Qie}%
  \BibitemOpen
  \bibfield  {author} {\bibinfo {author} {\bibfnamefont {Z.}~\bibnamefont
  {Qie}}, \bibinfo {author} {\bibfnamefont {H.}~\bibnamefont {Alhassawi}},
  \bibinfo {author} {\bibfnamefont {F.}~\bibnamefont {Sun}}, \bibinfo {author}
  {\bibfnamefont {J.}~\bibnamefont {Gao}}, \bibinfo {author} {\bibfnamefont
  {G.}~\bibnamefont {Zhao}},\ and\ \bibinfo {author} {\bibfnamefont
  {X.}~\bibnamefont {Fan}},\ }\bibfield  {title} {\enquote {\bibinfo {title}
  {Characteristics and applications of micro fluidized beds ({MFBs})},}\
  }\href@noop {} {\bibfield  {journal} {\bibinfo  {journal} {Chem. Eng. J.}\
  }\textbf {\bibinfo {volume} {428}},\ \bibinfo {pages} {131330} (\bibinfo
  {year} {2022})}\BibitemShut {NoStop}%
\bibitem [{\citenamefont {Schreiber}\ \emph {et~al.}(2002)\citenamefont
  {Schreiber}, \citenamefont {Vogt}, \citenamefont {Werther},\ and\
  \citenamefont {Brunner}}]{Schreiber}%
  \BibitemOpen
  \bibfield  {author} {\bibinfo {author} {\bibfnamefont {R.}~\bibnamefont
  {Schreiber}}, \bibinfo {author} {\bibfnamefont {C.}~\bibnamefont {Vogt}},
  \bibinfo {author} {\bibfnamefont {J.}~\bibnamefont {Werther}},\ and\ \bibinfo
  {author} {\bibfnamefont {G.}~\bibnamefont {Brunner}},\ }\bibfield  {title}
  {\enquote {\bibinfo {title} {Fluidized bed coating at supercritical fluid
  conditions},}\ }\href@noop {} {\bibfield  {journal} {\bibinfo  {journal} {J.
  Supercrit. Fluid.}\ }\textbf {\bibinfo {volume} {24}},\ \bibinfo {pages}
  {137--151} (\bibinfo {year} {2002})}\BibitemShut {NoStop}%
\bibitem [{\citenamefont {Rodr\'iguez-Rojo}, \citenamefont {Marienfeld},\ and\
  \citenamefont {Cocero}(2008)}]{Rodriguez}%
  \BibitemOpen
  \bibfield  {author} {\bibinfo {author} {\bibfnamefont {S.}~\bibnamefont
  {Rodr\'iguez-Rojo}}, \bibinfo {author} {\bibfnamefont {J.}~\bibnamefont
  {Marienfeld}},\ and\ \bibinfo {author} {\bibfnamefont {M.}~\bibnamefont
  {Cocero}},\ }\bibfield  {title} {\enquote {\bibinfo {title} {Ress process in
  coating applications in a high pressure fluidized bed environment: Bottom and
  top spray experiments},}\ }\href@noop {} {\bibfield  {journal} {\bibinfo
  {journal} {Chem. Eng. J.}\ }\textbf {\bibinfo {volume} {144}},\ \bibinfo
  {pages} {531--539} (\bibinfo {year} {2008})}\BibitemShut {NoStop}%
\bibitem [{\citenamefont {Jia}\ \emph {et~al.}(2017)\citenamefont {Jia},
  \citenamefont {Dufour}, \citenamefont {{Le Brech}}, \citenamefont {Authier},\
  and\ \citenamefont {Mauviel}}]{Jia}%
  \BibitemOpen
  \bibfield  {author} {\bibinfo {author} {\bibfnamefont {L.}~\bibnamefont
  {Jia}}, \bibinfo {author} {\bibfnamefont {A.}~\bibnamefont {Dufour}},
  \bibinfo {author} {\bibfnamefont {Y.}~\bibnamefont {{Le Brech}}}, \bibinfo
  {author} {\bibfnamefont {O.}~\bibnamefont {Authier}},\ and\ \bibinfo {author}
  {\bibfnamefont {G.}~\bibnamefont {Mauviel}},\ }\bibfield  {title} {\enquote
  {\bibinfo {title} {On-line analysis of primary tars from biomass pyrolysis by
  single photoionization mass spectrometry: {E}xperiments and detailed
  modelling},}\ }\href@noop {} {\bibfield  {journal} {\bibinfo  {journal}
  {Chem. Eng. J.}\ }\textbf {\bibinfo {volume} {313}},\ \bibinfo {pages}
  {270--282} (\bibinfo {year} {2017})}\BibitemShut {NoStop}%
\bibitem [{\citenamefont {Gao}\ \emph {et~al.}(2017)\citenamefont {Gao},
  \citenamefont {Farahani}, \citenamefont {Jamil}, \citenamefont {Siddiqui},
  \citenamefont {Siddiqui}, \citenamefont {Imran},\ and\ \citenamefont
  {Rezaee-Manesh}}]{Gao}%
  \BibitemOpen
  \bibfield  {author} {\bibinfo {author} {\bibfnamefont {W.}~\bibnamefont
  {Gao}}, \bibinfo {author} {\bibfnamefont {M.~R.}\ \bibnamefont {Farahani}},
  \bibinfo {author} {\bibfnamefont {M.~K.}\ \bibnamefont {Jamil}}, \bibinfo
  {author} {\bibfnamefont {M.~K.}\ \bibnamefont {Siddiqui}}, \bibinfo {author}
  {\bibfnamefont {H.~M.~A.}\ \bibnamefont {Siddiqui}}, \bibinfo {author}
  {\bibfnamefont {M.}~\bibnamefont {Imran}},\ and\ \bibinfo {author}
  {\bibfnamefont {R.}~\bibnamefont {Rezaee-Manesh}},\ }\bibfield  {title}
  {\enquote {\bibinfo {title} {Kinetic modeling of pyrolysis of three iranian
  waste oils in a micro-fluidized bed},}\ }\href@noop {} {\bibfield  {journal}
  {\bibinfo  {journal} {Pet. Sci. Technol.}\ }\textbf {\bibinfo {volume}
  {35}},\ \bibinfo {pages} {183--189} (\bibinfo {year} {2017})}\BibitemShut
  {NoStop}%
\bibitem [{\citenamefont {Mao}\ \emph {et~al.}(2015)\citenamefont {Mao},
  \citenamefont {Dong}, \citenamefont {Dong}, \citenamefont {Liu},
  \citenamefont {Chang}, \citenamefont {Yang}, \citenamefont {Lv},\ and\
  \citenamefont {Fan}}]{Mao}%
  \BibitemOpen
  \bibfield  {author} {\bibinfo {author} {\bibfnamefont {Y.}~\bibnamefont
  {Mao}}, \bibinfo {author} {\bibfnamefont {L.}~\bibnamefont {Dong}}, \bibinfo
  {author} {\bibfnamefont {Y.}~\bibnamefont {Dong}}, \bibinfo {author}
  {\bibfnamefont {W.}~\bibnamefont {Liu}}, \bibinfo {author} {\bibfnamefont
  {J.}~\bibnamefont {Chang}}, \bibinfo {author} {\bibfnamefont
  {S.}~\bibnamefont {Yang}}, \bibinfo {author} {\bibfnamefont {Z.}~\bibnamefont
  {Lv}},\ and\ \bibinfo {author} {\bibfnamefont {P.}~\bibnamefont {Fan}},\
  }\bibfield  {title} {\enquote {\bibinfo {title} {Fast co-pyrolysis of biomass
  and lignite in a micro fluidized bed reactor analyzer},}\ }\href@noop {}
  {\bibfield  {journal} {\bibinfo  {journal} {Bioresource Technol.}\ }\textbf
  {\bibinfo {volume} {181}},\ \bibinfo {pages} {155--162} (\bibinfo {year}
  {2015})}\BibitemShut {NoStop}%
\bibitem [{\citenamefont {Yu}\ \emph {et~al.}(2011)\citenamefont {Yu},
  \citenamefont {Yao}, \citenamefont {Zeng}, \citenamefont {Geng},
  \citenamefont {Dong}, \citenamefont {Wang}, \citenamefont {Gao},\ and\
  \citenamefont {Xu}}]{Yu}%
  \BibitemOpen
  \bibfield  {author} {\bibinfo {author} {\bibfnamefont {J.}~\bibnamefont
  {Yu}}, \bibinfo {author} {\bibfnamefont {C.}~\bibnamefont {Yao}}, \bibinfo
  {author} {\bibfnamefont {X.}~\bibnamefont {Zeng}}, \bibinfo {author}
  {\bibfnamefont {S.}~\bibnamefont {Geng}}, \bibinfo {author} {\bibfnamefont
  {L.}~\bibnamefont {Dong}}, \bibinfo {author} {\bibfnamefont {Y.}~\bibnamefont
  {Wang}}, \bibinfo {author} {\bibfnamefont {S.}~\bibnamefont {Gao}},\ and\
  \bibinfo {author} {\bibfnamefont {G.}~\bibnamefont {Xu}},\ }\bibfield
  {title} {\enquote {\bibinfo {title} {Biomass pyrolysis in a micro-fluidized
  bed reactor: Characterization and kinetics},}\ }\href@noop {} {\bibfield
  {journal} {\bibinfo  {journal} {Chem. Eng. J.}\ }\textbf {\bibinfo {volume}
  {168}},\ \bibinfo {pages} {839--847} (\bibinfo {year} {2011})}\BibitemShut
  {NoStop}%
\bibitem [{\citenamefont {Boffito}\ \emph {et~al.}(2014)\citenamefont
  {Boffito}, \citenamefont {Neagoe}, \citenamefont {Edake}, \citenamefont
  {Pastor-Ramirez},\ and\ \citenamefont {Patience}}]{Boffito}%
  \BibitemOpen
  \bibfield  {author} {\bibinfo {author} {\bibfnamefont {D.}~\bibnamefont
  {Boffito}}, \bibinfo {author} {\bibfnamefont {C.}~\bibnamefont {Neagoe}},
  \bibinfo {author} {\bibfnamefont {M.}~\bibnamefont {Edake}}, \bibinfo
  {author} {\bibfnamefont {B.}~\bibnamefont {Pastor-Ramirez}},\ and\ \bibinfo
  {author} {\bibfnamefont {G.}~\bibnamefont {Patience}},\ }\bibfield  {title}
  {\enquote {\bibinfo {title} {Biofuel synthesis in a capillary fluidized
  bed},}\ }\href@noop {} {\bibfield  {journal} {\bibinfo  {journal} {Catal.
  Today}\ }\textbf {\bibinfo {volume} {237}},\ \bibinfo {pages} {13--17}
  (\bibinfo {year} {2014})}\BibitemShut {NoStop}%
\bibitem [{\citenamefont {Guo}\ \emph {et~al.}(2016)\citenamefont {Guo},
  \citenamefont {Dong}, \citenamefont {Fan}, \citenamefont {Lv}, \citenamefont
  {Yang},\ and\ \citenamefont {Dong}}]{Guo2}%
  \BibitemOpen
  \bibfield  {author} {\bibinfo {author} {\bibfnamefont {F.}~\bibnamefont
  {Guo}}, \bibinfo {author} {\bibfnamefont {Y.}~\bibnamefont {Dong}}, \bibinfo
  {author} {\bibfnamefont {P.}~\bibnamefont {Fan}}, \bibinfo {author}
  {\bibfnamefont {Z.}~\bibnamefont {Lv}}, \bibinfo {author} {\bibfnamefont
  {S.}~\bibnamefont {Yang}},\ and\ \bibinfo {author} {\bibfnamefont
  {L.}~\bibnamefont {Dong}},\ }\bibfield  {title} {\enquote {\bibinfo {title}
  {Catalytic decomposition of biomass tar compound by calcined coal gangue: {A}
  kinetic study},}\ }\href@noop {} {\bibfield  {journal} {\bibinfo  {journal}
  {Int. J. Hydrogen Energ.}\ }\textbf {\bibinfo {volume} {41}},\ \bibinfo
  {pages} {13380--13389} (\bibinfo {year} {2016})}\BibitemShut {NoStop}%
\bibitem [{\citenamefont {Zeng}\ \emph {et~al.}(2014)\citenamefont {Zeng},
  \citenamefont {Wang}, \citenamefont {Wang}, \citenamefont {Li}, \citenamefont
  {Yu},\ and\ \citenamefont {Xu}}]{Zeng}%
  \BibitemOpen
  \bibfield  {author} {\bibinfo {author} {\bibfnamefont {X.}~\bibnamefont
  {Zeng}}, \bibinfo {author} {\bibfnamefont {F.}~\bibnamefont {Wang}}, \bibinfo
  {author} {\bibfnamefont {Y.}~\bibnamefont {Wang}}, \bibinfo {author}
  {\bibfnamefont {A.}~\bibnamefont {Li}}, \bibinfo {author} {\bibfnamefont
  {J.}~\bibnamefont {Yu}},\ and\ \bibinfo {author} {\bibfnamefont
  {G.}~\bibnamefont {Xu}},\ }\bibfield  {title} {\enquote {\bibinfo {title}
  {Characterization of char gasification in a micro fluidized bed reaction
  analyzer},}\ }\href@noop {} {\bibfield  {journal} {\bibinfo  {journal}
  {Energy \& Fuels}\ }\textbf {\bibinfo {volume} {28}},\ \bibinfo {pages}
  {1838--1845} (\bibinfo {year} {2014})}\BibitemShut {NoStop}%
\bibitem [{\citenamefont {Zhang}\ \emph {et~al.}(2015)\citenamefont {Zhang},
  \citenamefont {Yao}, \citenamefont {Gao}, \citenamefont {Sun},\ and\
  \citenamefont {Xu}}]{Zhang2}%
  \BibitemOpen
  \bibfield  {author} {\bibinfo {author} {\bibfnamefont {Y.}~\bibnamefont
  {Zhang}}, \bibinfo {author} {\bibfnamefont {M.}~\bibnamefont {Yao}}, \bibinfo
  {author} {\bibfnamefont {S.}~\bibnamefont {Gao}}, \bibinfo {author}
  {\bibfnamefont {G.}~\bibnamefont {Sun}},\ and\ \bibinfo {author}
  {\bibfnamefont {G.}~\bibnamefont {Xu}},\ }\bibfield  {title} {\enquote
  {\bibinfo {title} {Reactivity and kinetics for steam gasification of
  petroleum coke blended with black liquor in a micro fluidized bed},}\
  }\href@noop {} {\bibfield  {journal} {\bibinfo  {journal} {Applied Energy}\
  }\textbf {\bibinfo {volume} {160}},\ \bibinfo {pages} {820--828} (\bibinfo
  {year} {2015})}\BibitemShut {NoStop}%
\bibitem [{\citenamefont {Cortazar}\ \emph {et~al.}(2020)\citenamefont
  {Cortazar}, \citenamefont {Lopez}, \citenamefont {Alvarez}, \citenamefont
  {Arregi}, \citenamefont {Amutio}, \citenamefont {Bilbao},\ and\ \citenamefont
  {Olazar}}]{Cortazar}%
  \BibitemOpen
  \bibfield  {author} {\bibinfo {author} {\bibfnamefont {M.}~\bibnamefont
  {Cortazar}}, \bibinfo {author} {\bibfnamefont {G.}~\bibnamefont {Lopez}},
  \bibinfo {author} {\bibfnamefont {J.}~\bibnamefont {Alvarez}}, \bibinfo
  {author} {\bibfnamefont {A.}~\bibnamefont {Arregi}}, \bibinfo {author}
  {\bibfnamefont {M.}~\bibnamefont {Amutio}}, \bibinfo {author} {\bibfnamefont
  {J.}~\bibnamefont {Bilbao}},\ and\ \bibinfo {author} {\bibfnamefont
  {M.}~\bibnamefont {Olazar}},\ }\bibfield  {title} {\enquote {\bibinfo {title}
  {Experimental study and modeling of biomass char gasification kinetics in a
  novel thermogravimetric flow reactor},}\ }\href@noop {} {\bibfield  {journal}
  {\bibinfo  {journal} {Chem. Eng. J.}\ }\textbf {\bibinfo {volume} {396}},\
  \bibinfo {pages} {125200} (\bibinfo {year} {2020})}\BibitemShut {NoStop}%
\bibitem [{\citenamefont {Fang}, \citenamefont {Li},\ and\ \citenamefont
  {Cai}(2009)}]{Fang}%
  \BibitemOpen
  \bibfield  {author} {\bibinfo {author} {\bibfnamefont {F.}~\bibnamefont
  {Fang}}, \bibinfo {author} {\bibfnamefont {Z.-S.}\ \bibnamefont {Li}},\ and\
  \bibinfo {author} {\bibfnamefont {N.-S.}\ \bibnamefont {Cai}},\ }\bibfield
  {title} {\enquote {\bibinfo {title} {Experiment and modeling of co2 capture
  from flue gases at high temperature in a fluidized bed reactor with ca-based
  sorbents},}\ }\href@noop {} {\bibfield  {journal} {\bibinfo  {journal}
  {Energy \& Fuels}\ }\textbf {\bibinfo {volume} {23}},\ \bibinfo {pages}
  {207--216} (\bibinfo {year} {2009})}\BibitemShut {NoStop}%
\bibitem [{\citenamefont {Shen}\ \emph {et~al.}(2019)\citenamefont {Shen},
  \citenamefont {Zhu}, \citenamefont {Yan},\ and\ \citenamefont {Shen}}]{Shen}%
  \BibitemOpen
  \bibfield  {author} {\bibinfo {author} {\bibfnamefont {T.}~\bibnamefont
  {Shen}}, \bibinfo {author} {\bibfnamefont {X.}~\bibnamefont {Zhu}}, \bibinfo
  {author} {\bibfnamefont {J.}~\bibnamefont {Yan}},\ and\ \bibinfo {author}
  {\bibfnamefont {L.}~\bibnamefont {Shen}},\ }\bibfield  {title} {\enquote
  {\bibinfo {title} {Design of micro interconnected fluidized bed for oxygen
  carrier evaluation},}\ }\href@noop {} {\bibfield  {journal} {\bibinfo
  {journal} {Int. J. Greenh. Gas Con.}\ }\textbf {\bibinfo {volume} {90}},\
  \bibinfo {pages} {102806} (\bibinfo {year} {2019})}\BibitemShut {NoStop}%
\bibitem [{\citenamefont {Wu}, \citenamefont {Chang},\ and\ \citenamefont
  {Chang}(2007)}]{Wu}%
  \BibitemOpen
  \bibfield  {author} {\bibinfo {author} {\bibfnamefont {K.-J.}\ \bibnamefont
  {Wu}}, \bibinfo {author} {\bibfnamefont {C.-F.}\ \bibnamefont {Chang}},\ and\
  \bibinfo {author} {\bibfnamefont {J.-S.}\ \bibnamefont {Chang}},\ }\bibfield
  {title} {\enquote {\bibinfo {title} {Simultaneous production of biohydrogen
  and bioethanol with fluidized-bed and packed-bed bioreactors containing
  immobilized anaerobic sludge},}\ }\href@noop {} {\bibfield  {journal}
  {\bibinfo  {journal} {Process Biochem.}\ }\textbf {\bibinfo {volume} {42}},\
  \bibinfo {pages} {1165--1171} (\bibinfo {year} {2007})}\BibitemShut {NoStop}%
\bibitem [{\citenamefont {Liu}, \citenamefont {Ren},\ and\ \citenamefont
  {Yao}(2010)}]{Liu4}%
  \BibitemOpen
  \bibfield  {author} {\bibinfo {author} {\bibfnamefont {J.}~\bibnamefont
  {Liu}}, \bibinfo {author} {\bibfnamefont {Y.}~\bibnamefont {Ren}},\ and\
  \bibinfo {author} {\bibfnamefont {S.}~\bibnamefont {Yao}},\ }\bibfield
  {title} {\enquote {\bibinfo {title} {Repeated-batch cultivation of
  encapsulated monascus purpureus by polyelectrolyte complex for natural
  pigment production},}\ }\href@noop {} {\bibfield  {journal} {\bibinfo
  {journal} {Chinese J. Chem. Eng.}\ }\textbf {\bibinfo {volume} {18}},\
  \bibinfo {pages} {1013--1017} (\bibinfo {year} {2010})}\BibitemShut {NoStop}%
\bibitem [{\citenamefont {Pereiro}\ \emph {et~al.}(2017)\citenamefont
  {Pereiro}, \citenamefont {Bendali}, \citenamefont {Tabnaoui}, \citenamefont
  {Alexandre}, \citenamefont {Srbova}, \citenamefont {Bilkova}, \citenamefont
  {Deegan}, \citenamefont {Joshi}, \citenamefont {Viovy}, \citenamefont
  {Malaquin}, \citenamefont {Dupuy},\ and\ \citenamefont {Descroix}}]{Pereiro}%
  \BibitemOpen
  \bibfield  {author} {\bibinfo {author} {\bibfnamefont {I.}~\bibnamefont
  {Pereiro}}, \bibinfo {author} {\bibfnamefont {A.}~\bibnamefont {Bendali}},
  \bibinfo {author} {\bibfnamefont {S.}~\bibnamefont {Tabnaoui}}, \bibinfo
  {author} {\bibfnamefont {L.}~\bibnamefont {Alexandre}}, \bibinfo {author}
  {\bibfnamefont {J.}~\bibnamefont {Srbova}}, \bibinfo {author} {\bibfnamefont
  {Z.}~\bibnamefont {Bilkova}}, \bibinfo {author} {\bibfnamefont
  {S.}~\bibnamefont {Deegan}}, \bibinfo {author} {\bibfnamefont
  {L.}~\bibnamefont {Joshi}}, \bibinfo {author} {\bibfnamefont {J.-L.}\
  \bibnamefont {Viovy}}, \bibinfo {author} {\bibfnamefont {L.}~\bibnamefont
  {Malaquin}}, \bibinfo {author} {\bibfnamefont {B.}~\bibnamefont {Dupuy}},\
  and\ \bibinfo {author} {\bibfnamefont {S.}~\bibnamefont {Descroix}},\
  }\bibfield  {title} {\enquote {\bibinfo {title} {A new microfluidic approach
  for the one-step capture{,} amplification and label-free quantification of
  bacteria from raw samples},}\ }\href@noop {} {\bibfield  {journal} {\bibinfo
  {journal} {Chem. Sci.}\ }\textbf {\bibinfo {volume} {8}},\ \bibinfo {pages}
  {1329--1336} (\bibinfo {year} {2017})}\BibitemShut {NoStop}%
\bibitem [{\citenamefont {Kuyukina}\ \emph {et~al.}(2009)\citenamefont
  {Kuyukina}, \citenamefont {Ivshina}, \citenamefont {Serebrennikova},
  \citenamefont {Krivorutchko}, \citenamefont {Podorozhko}, \citenamefont
  {Ivanov},\ and\ \citenamefont {Lozinsky}}]{Kuyukina}%
  \BibitemOpen
  \bibfield  {author} {\bibinfo {author} {\bibfnamefont {M.~S.}\ \bibnamefont
  {Kuyukina}}, \bibinfo {author} {\bibfnamefont {I.~B.}\ \bibnamefont
  {Ivshina}}, \bibinfo {author} {\bibfnamefont {M.~K.}\ \bibnamefont
  {Serebrennikova}}, \bibinfo {author} {\bibfnamefont {A.~B.}\ \bibnamefont
  {Krivorutchko}}, \bibinfo {author} {\bibfnamefont {E.~A.}\ \bibnamefont
  {Podorozhko}}, \bibinfo {author} {\bibfnamefont {R.~V.}\ \bibnamefont
  {Ivanov}},\ and\ \bibinfo {author} {\bibfnamefont {V.~I.}\ \bibnamefont
  {Lozinsky}},\ }\bibfield  {title} {\enquote {\bibinfo {title}
  {Petroleum-contaminated water treatment in a fluidized-bed bioreactor with
  immobilized rhodococcus cells},}\ }\href@noop {} {\bibfield  {journal}
  {\bibinfo  {journal} {Int. Biodeter. Biodegr.}\ }\textbf {\bibinfo {volume}
  {63}},\ \bibinfo {pages} {427--432} (\bibinfo {year} {2009})}\BibitemShut
  {NoStop}%
\bibitem [{\citenamefont {Kwak}\ \emph {et~al.}(2020)\citenamefont {Kwak},
  \citenamefont {Rout}, \citenamefont {Lee},\ and\ \citenamefont {Bae}}]{Kwak}%
  \BibitemOpen
  \bibfield  {author} {\bibinfo {author} {\bibfnamefont {W.}~\bibnamefont
  {Kwak}}, \bibinfo {author} {\bibfnamefont {P.~R.}\ \bibnamefont {Rout}},
  \bibinfo {author} {\bibfnamefont {E.}~\bibnamefont {Lee}},\ and\ \bibinfo
  {author} {\bibfnamefont {J.}~\bibnamefont {Bae}},\ }\bibfield  {title}
  {\enquote {\bibinfo {title} {Influence of hydraulic retention time and
  temperature on the performance of an anaerobic ammonium oxidation fluidized
  bed membrane bioreactor for low-strength ammonia wastewater treatment},}\
  }\href@noop {} {\bibfield  {journal} {\bibinfo  {journal} {Chem. Eng. J.}\
  }\textbf {\bibinfo {volume} {386}},\ \bibinfo {pages} {123992} (\bibinfo
  {year} {2020})}\BibitemShut {NoStop}%
\bibitem [{\citenamefont {Kornblum}\ and\ \citenamefont
  {Hirschorn}(1970)}]{Kornblum}%
  \BibitemOpen
  \bibfield  {author} {\bibinfo {author} {\bibfnamefont {S.~S.}\ \bibnamefont
  {Kornblum}}\ and\ \bibinfo {author} {\bibfnamefont {J.~O.}\ \bibnamefont
  {Hirschorn}},\ }\bibfield  {title} {\enquote {\bibinfo {title} {Dissolution
  of poorly water-soluble drugs i: Some physical parameters related to method
  of micronization and tablet manufacture of a quinazolinone compound},}\
  }\href@noop {} {\bibfield  {journal} {\bibinfo  {journal} {J. Pharm. Sci.}\
  }\textbf {\bibinfo {volume} {59}},\ \bibinfo {pages} {606--609} (\bibinfo
  {year} {1970})}\BibitemShut {NoStop}%
\bibitem [{\citenamefont {Katstra}\ \emph {et~al.}(2000)\citenamefont
  {Katstra}, \citenamefont {Palazzolo}, \citenamefont {Rowe}, \citenamefont
  {Giritlioglu}, \citenamefont {Teung},\ and\ \citenamefont {Cima}}]{Katstra}%
  \BibitemOpen
  \bibfield  {author} {\bibinfo {author} {\bibfnamefont {W.}~\bibnamefont
  {Katstra}}, \bibinfo {author} {\bibfnamefont {R.}~\bibnamefont {Palazzolo}},
  \bibinfo {author} {\bibfnamefont {C.}~\bibnamefont {Rowe}}, \bibinfo {author}
  {\bibfnamefont {B.}~\bibnamefont {Giritlioglu}}, \bibinfo {author}
  {\bibfnamefont {P.}~\bibnamefont {Teung}},\ and\ \bibinfo {author}
  {\bibfnamefont {M.}~\bibnamefont {Cima}},\ }\bibfield  {title} {\enquote
  {\bibinfo {title} {Oral dosage forms fabricated by three dimensional
  printing™},}\ }\href@noop {} {\bibfield  {journal} {\bibinfo  {journal} {J.
  Control. Release}\ }\textbf {\bibinfo {volume} {66}},\ \bibinfo {pages}
  {1--9} (\bibinfo {year} {2000})}\BibitemShut {NoStop}%
\bibitem [{\citenamefont {Fung}\ and\ \citenamefont {Ng}(2003)}]{Fung}%
  \BibitemOpen
  \bibfield  {author} {\bibinfo {author} {\bibfnamefont {K.~Y.}\ \bibnamefont
  {Fung}}\ and\ \bibinfo {author} {\bibfnamefont {K.~M.}\ \bibnamefont {Ng}},\
  }\bibfield  {title} {\enquote {\bibinfo {title} {Product-centered processing:
  Pharmaceutical tablets and capsules},}\ }\href@noop {} {\bibfield  {journal}
  {\bibinfo  {journal} {AIChE Journal}\ }\textbf {\bibinfo {volume} {49}},\
  \bibinfo {pages} {1193--1215} (\bibinfo {year} {2003})}\BibitemShut {NoStop}%
\bibitem [{\citenamefont {Ervasti}\ \emph {et~al.}(2015)\citenamefont
  {Ervasti}, \citenamefont {Simonaho}, \citenamefont {Ketolainen},
  \citenamefont {Forsberg}, \citenamefont {Fransson}, \citenamefont
  {Wikström}, \citenamefont {Folestad}, \citenamefont {Lakio}, \citenamefont
  {Tajarobi},\ and\ \citenamefont {Abrahms\'en-Alami}}]{Ervasti}%
  \BibitemOpen
  \bibfield  {author} {\bibinfo {author} {\bibfnamefont {T.}~\bibnamefont
  {Ervasti}}, \bibinfo {author} {\bibfnamefont {S.-P.}\ \bibnamefont
  {Simonaho}}, \bibinfo {author} {\bibfnamefont {J.}~\bibnamefont
  {Ketolainen}}, \bibinfo {author} {\bibfnamefont {P.}~\bibnamefont
  {Forsberg}}, \bibinfo {author} {\bibfnamefont {M.}~\bibnamefont {Fransson}},
  \bibinfo {author} {\bibfnamefont {H.}~\bibnamefont {Wikström}}, \bibinfo
  {author} {\bibfnamefont {S.}~\bibnamefont {Folestad}}, \bibinfo {author}
  {\bibfnamefont {S.}~\bibnamefont {Lakio}}, \bibinfo {author} {\bibfnamefont
  {P.}~\bibnamefont {Tajarobi}},\ and\ \bibinfo {author} {\bibfnamefont
  {S.}~\bibnamefont {Abrahms\'en-Alami}},\ }\bibfield  {title} {\enquote
  {\bibinfo {title} {Continuous manufacturing of extended release tablets via
  powder mixing and direct compression},}\ }\href@noop {} {\bibfield  {journal}
  {\bibinfo  {journal} {Internat. J. Pharmaceut.}\ }\textbf {\bibinfo {volume}
  {495}},\ \bibinfo {pages} {290--301} (\bibinfo {year} {2015})}\BibitemShut
  {NoStop}%
\bibitem [{\citenamefont {Azad}\ \emph {et~al.}(2018)\citenamefont {Azad},
  \citenamefont {Osorio}, \citenamefont {Brancazio}, \citenamefont
  {Hammersmith}, \citenamefont {Klee}, \citenamefont {Rapp},\ and\
  \citenamefont {Myerson}}]{Azad}%
  \BibitemOpen
  \bibfield  {author} {\bibinfo {author} {\bibfnamefont {M.~A.}\ \bibnamefont
  {Azad}}, \bibinfo {author} {\bibfnamefont {J.~G.}\ \bibnamefont {Osorio}},
  \bibinfo {author} {\bibfnamefont {D.}~\bibnamefont {Brancazio}}, \bibinfo
  {author} {\bibfnamefont {G.}~\bibnamefont {Hammersmith}}, \bibinfo {author}
  {\bibfnamefont {D.~M.}\ \bibnamefont {Klee}}, \bibinfo {author}
  {\bibfnamefont {K.}~\bibnamefont {Rapp}},\ and\ \bibinfo {author}
  {\bibfnamefont {A.}~\bibnamefont {Myerson}},\ }\bibfield  {title} {\enquote
  {\bibinfo {title} {A compact, portable, re-configurable, and automated system
  for on-demand pharmaceutical tablet manufacturing},}\ }\href@noop {}
  {\bibfield  {journal} {\bibinfo  {journal} {Internat. J. Pharmaceut.}\
  }\textbf {\bibinfo {volume} {539}},\ \bibinfo {pages} {157--164} (\bibinfo
  {year} {2018})}\BibitemShut {NoStop}%
\bibitem [{\citenamefont {Azad}\ \emph {et~al.}(2019)\citenamefont {Azad},
  \citenamefont {Osorio}, \citenamefont {Wang}, \citenamefont {Klee},
  \citenamefont {Eccles}, \citenamefont {Grela}, \citenamefont {Sloan},
  \citenamefont {Hammersmith}, \citenamefont {Rapp}, \citenamefont
  {Brancazio},\ and\ \citenamefont {Myerson}}]{Azad2}%
  \BibitemOpen
  \bibfield  {author} {\bibinfo {author} {\bibfnamefont {M.~A.}\ \bibnamefont
  {Azad}}, \bibinfo {author} {\bibfnamefont {J.~G.}\ \bibnamefont {Osorio}},
  \bibinfo {author} {\bibfnamefont {A.}~\bibnamefont {Wang}}, \bibinfo {author}
  {\bibfnamefont {D.~M.}\ \bibnamefont {Klee}}, \bibinfo {author}
  {\bibfnamefont {M.~E.}\ \bibnamefont {Eccles}}, \bibinfo {author}
  {\bibfnamefont {E.}~\bibnamefont {Grela}}, \bibinfo {author} {\bibfnamefont
  {R.}~\bibnamefont {Sloan}}, \bibinfo {author} {\bibfnamefont
  {G.}~\bibnamefont {Hammersmith}}, \bibinfo {author} {\bibfnamefont
  {K.}~\bibnamefont {Rapp}}, \bibinfo {author} {\bibfnamefont {D.}~\bibnamefont
  {Brancazio}},\ and\ \bibinfo {author} {\bibfnamefont {A.~S.}\ \bibnamefont
  {Myerson}},\ }\bibfield  {title} {\enquote {\bibinfo {title} {On-demand
  manufacturing of direct compressible tablets: {C}an formulation be
  simplified?}}\ }\href@noop {} {\bibfield  {journal} {\bibinfo  {journal}
  {Pharm Res}\ }\textbf {\bibinfo {volume} {36}},\ \bibinfo {pages} {167}
  (\bibinfo {year} {2019})}\BibitemShut {NoStop}%
\bibitem [{\citenamefont {Shi}\ \emph {et~al.}(2019)\citenamefont {Shi},
  \citenamefont {Tan}, \citenamefont {Nokhodchi},\ and\ \citenamefont
  {Maniruzzaman}}]{Shi}%
  \BibitemOpen
  \bibfield  {author} {\bibinfo {author} {\bibfnamefont {K.}~\bibnamefont
  {Shi}}, \bibinfo {author} {\bibfnamefont {D.~K.}\ \bibnamefont {Tan}},
  \bibinfo {author} {\bibfnamefont {A.}~\bibnamefont {Nokhodchi}},\ and\
  \bibinfo {author} {\bibfnamefont {M.}~\bibnamefont {Maniruzzaman}},\
  }\bibfield  {title} {\enquote {\bibinfo {title} {Drop-on-powder 3d printing
  of tablets with an anti-cancer drug, 5-fluorouracil},}\ }\href@noop {}
  {\bibfield  {journal} {\bibinfo  {journal} {Pharmaceutics}\ }\textbf
  {\bibinfo {volume} {11}},\ \bibinfo {pages} {150} (\bibinfo {year}
  {2019})}\BibitemShut {NoStop}%
\bibitem [{\citenamefont {Jiang}\ \emph {et~al.}(2006)\citenamefont {Jiang},
  \citenamefont {Joshi}, \citenamefont {Peek}, \citenamefont {Brandau},
  \citenamefont {Huang}, \citenamefont {Ferriter}, \citenamefont {Woodley},
  \citenamefont {Ford}, \citenamefont {Mar}, \citenamefont {Mikszta},
  \citenamefont {Hwang}, \citenamefont {Ulrich}, \citenamefont {Harvey},
  \citenamefont {Middaugh},\ and\ \citenamefont {Sullivan}}]{Jiang}%
  \BibitemOpen
  \bibfield  {author} {\bibinfo {author} {\bibfnamefont {G.}~\bibnamefont
  {Jiang}}, \bibinfo {author} {\bibfnamefont {S.~B.}\ \bibnamefont {Joshi}},
  \bibinfo {author} {\bibfnamefont {L.~J.}\ \bibnamefont {Peek}}, \bibinfo
  {author} {\bibfnamefont {D.~T.}\ \bibnamefont {Brandau}}, \bibinfo {author}
  {\bibfnamefont {J.}~\bibnamefont {Huang}}, \bibinfo {author} {\bibfnamefont
  {M.~S.}\ \bibnamefont {Ferriter}}, \bibinfo {author} {\bibfnamefont {W.~D.}\
  \bibnamefont {Woodley}}, \bibinfo {author} {\bibfnamefont {B.~M.}\
  \bibnamefont {Ford}}, \bibinfo {author} {\bibfnamefont {K.~D.}\ \bibnamefont
  {Mar}}, \bibinfo {author} {\bibfnamefont {J.~A.}\ \bibnamefont {Mikszta}},
  \bibinfo {author} {\bibfnamefont {C.}~\bibnamefont {Hwang}}, \bibinfo
  {author} {\bibfnamefont {R.}~\bibnamefont {Ulrich}}, \bibinfo {author}
  {\bibfnamefont {N.~G.}\ \bibnamefont {Harvey}}, \bibinfo {author}
  {\bibfnamefont {C.}~\bibnamefont {Middaugh}},\ and\ \bibinfo {author}
  {\bibfnamefont {V.~J.}\ \bibnamefont {Sullivan}},\ }\bibfield  {title}
  {\enquote {\bibinfo {title} {Anthrax vaccine powder formulations for nasal
  mucosal delivery},}\ }\href@noop {} {\bibfield  {journal} {\bibinfo
  {journal} {J. Pharm. Sci.}\ }\textbf {\bibinfo {volume} {95}},\ \bibinfo
  {pages} {80--96} (\bibinfo {year} {2006})}\BibitemShut {NoStop}%
\bibitem [{\citenamefont {Huang}\ \emph {et~al.}(2004)\citenamefont {Huang},
  \citenamefont {Garmise}, \citenamefont {Crowder}, \citenamefont {Mar},
  \citenamefont {Hwang}, \citenamefont {Hickey}, \citenamefont {Mikszta},\ and\
  \citenamefont {Sullivan}}]{Huang}%
  \BibitemOpen
  \bibfield  {author} {\bibinfo {author} {\bibfnamefont {J.}~\bibnamefont
  {Huang}}, \bibinfo {author} {\bibfnamefont {R.~J.}\ \bibnamefont {Garmise}},
  \bibinfo {author} {\bibfnamefont {T.~M.}\ \bibnamefont {Crowder}}, \bibinfo
  {author} {\bibfnamefont {K.}~\bibnamefont {Mar}}, \bibinfo {author}
  {\bibfnamefont {C.~R.}\ \bibnamefont {Hwang}}, \bibinfo {author}
  {\bibfnamefont {A.~J.}\ \bibnamefont {Hickey}}, \bibinfo {author}
  {\bibfnamefont {J.~A.}\ \bibnamefont {Mikszta}},\ and\ \bibinfo {author}
  {\bibfnamefont {V.~J.}\ \bibnamefont {Sullivan}},\ }\bibfield  {title}
  {\enquote {\bibinfo {title} {A novel dry powder influenza vaccine and
  intranasal delivery technology: induction of systemic and mucosal immune
  responses in rats},}\ }\href@noop {} {\bibfield  {journal} {\bibinfo
  {journal} {Vaccine}\ }\textbf {\bibinfo {volume} {23}},\ \bibinfo {pages}
  {794--801} (\bibinfo {year} {2004})}\BibitemShut {NoStop}%
\bibitem [{\citenamefont {Gomez}\ \emph {et~al.}(2021)\citenamefont {Gomez},
  \citenamefont {McCollum}, \citenamefont {Wang}, \citenamefont {Bachchhav},
  \citenamefont {Tetreau}, \citenamefont {Gerhardt}, \citenamefont {Press},
  \citenamefont {Kramer}, \citenamefont {Fox},\ and\ \citenamefont
  {Vehring}}]{Gomez}%
  \BibitemOpen
  \bibfield  {author} {\bibinfo {author} {\bibfnamefont {M.}~\bibnamefont
  {Gomez}}, \bibinfo {author} {\bibfnamefont {J.}~\bibnamefont {McCollum}},
  \bibinfo {author} {\bibfnamefont {H.}~\bibnamefont {Wang}}, \bibinfo {author}
  {\bibfnamefont {S.}~\bibnamefont {Bachchhav}}, \bibinfo {author}
  {\bibfnamefont {I.}~\bibnamefont {Tetreau}}, \bibinfo {author} {\bibfnamefont
  {A.}~\bibnamefont {Gerhardt}}, \bibinfo {author} {\bibfnamefont
  {C.}~\bibnamefont {Press}}, \bibinfo {author} {\bibfnamefont {R.~M.}\
  \bibnamefont {Kramer}}, \bibinfo {author} {\bibfnamefont {C.~B.}\
  \bibnamefont {Fox}},\ and\ \bibinfo {author} {\bibfnamefont {R.}~\bibnamefont
  {Vehring}},\ }\bibfield  {title} {\enquote {\bibinfo {title} {Evaluation of
  the stability of a spray-dried tuberculosis vaccine candidate designed for
  dry powder respiratory delivery},}\ }\href@noop {} {\bibfield  {journal}
  {\bibinfo  {journal} {Vaccine}\ }\textbf {\bibinfo {volume} {39}},\ \bibinfo
  {pages} {5025--5036} (\bibinfo {year} {2021})}\BibitemShut {NoStop}%
\bibitem [{\citenamefont {Heida}, \citenamefont {Hinrichs},\ and\ \citenamefont
  {Frijlink}(2022)}]{Heida}%
  \BibitemOpen
  \bibfield  {author} {\bibinfo {author} {\bibfnamefont {R.}~\bibnamefont
  {Heida}}, \bibinfo {author} {\bibfnamefont {W.~L.}\ \bibnamefont
  {Hinrichs}},\ and\ \bibinfo {author} {\bibfnamefont {H.~W.}\ \bibnamefont
  {Frijlink}},\ }\bibfield  {title} {\enquote {\bibinfo {title} {Inhaled
  vaccine delivery in the combat against respiratory viruses: a 2021 overview
  of recent developments and implications for covid-19},}\ }\href@noop {}
  {\bibfield  {journal} {\bibinfo  {journal} {Expert Rev. Vaccines}\ }\textbf
  {\bibinfo {volume} {21}},\ \bibinfo {pages} {957--974} (\bibinfo {year}
  {2022})}\BibitemShut {NoStop}%
\bibitem [{\citenamefont {Amorij}\ \emph {et~al.}(2008)\citenamefont {Amorij},
  \citenamefont {Huckriede}, \citenamefont {Wilschut}, \citenamefont
  {Frijlink},\ and\ \citenamefont {Hinrichs}}]{Amorij}%
  \BibitemOpen
  \bibfield  {author} {\bibinfo {author} {\bibfnamefont {J.~P.}\ \bibnamefont
  {Amorij}}, \bibinfo {author} {\bibfnamefont {A.}~\bibnamefont {Huckriede}},
  \bibinfo {author} {\bibfnamefont {J.}~\bibnamefont {Wilschut}}, \bibinfo
  {author} {\bibfnamefont {H.~W.}\ \bibnamefont {Frijlink}},\ and\ \bibinfo
  {author} {\bibfnamefont {W.~L.~J.}\ \bibnamefont {Hinrichs}},\ }\bibfield
  {title} {\enquote {\bibinfo {title} {Development of stable influenza vaccine
  powder formulations: {C}hallenges and possibilities},}\ }\href@noop {}
  {\bibfield  {journal} {\bibinfo  {journal} {Pharm. Res.}\ }\textbf {\bibinfo
  {volume} {25}},\ \bibinfo {pages} {1256--1273} (\bibinfo {year}
  {2008})}\BibitemShut {NoStop}%
\bibitem [{\citenamefont {Guo}, \citenamefont {Xu},\ and\ \citenamefont
  {Yue}(2009)}]{Guo}%
  \BibitemOpen
  \bibfield  {author} {\bibinfo {author} {\bibfnamefont {Q.~j.}\ \bibnamefont
  {Guo}}, \bibinfo {author} {\bibfnamefont {Y.}~\bibnamefont {Xu}},\ and\
  \bibinfo {author} {\bibfnamefont {X.}~\bibnamefont {Yue}},\ }\bibfield
  {title} {\enquote {\bibinfo {title} {Fluidization characteristics in
  micro-fluidized beds of various inner diameters},}\ }\href@noop {} {\bibfield
   {journal} {\bibinfo  {journal} {Chem. Eng. Technol.}\ }\textbf {\bibinfo
  {volume} {32}},\ \bibinfo {pages} {1992--1999} (\bibinfo {year}
  {2009})}\BibitemShut {NoStop}%
\bibitem [{\citenamefont {{do Nascimento}}, \citenamefont {Reay},\ and\
  \citenamefont {Zivkovic}(2016)}]{Nascimento}%
  \BibitemOpen
  \bibfield  {author} {\bibinfo {author} {\bibfnamefont {O.~L.}\ \bibnamefont
  {{do Nascimento}}}, \bibinfo {author} {\bibfnamefont {D.~A.}\ \bibnamefont
  {Reay}},\ and\ \bibinfo {author} {\bibfnamefont {V.}~\bibnamefont
  {Zivkovic}},\ }\bibfield  {title} {\enquote {\bibinfo {title} {Influence of
  surface forces and wall effects on the minimum fluidization velocity of
  liquid-solid micro-fluidized beds},}\ }\href@noop {} {\bibfield  {journal}
  {\bibinfo  {journal} {Powder Technol.}\ }\textbf {\bibinfo {volume} {304}},\
  \bibinfo {pages} {55--62} (\bibinfo {year} {2016})}\BibitemShut {NoStop}%
\bibitem [{\citenamefont {Li}, \citenamefont {Liu},\ and\ \citenamefont
  {Li}(2018)}]{Li2}%
  \BibitemOpen
  \bibfield  {author} {\bibinfo {author} {\bibfnamefont {X.}~\bibnamefont
  {Li}}, \bibinfo {author} {\bibfnamefont {M.}~\bibnamefont {Liu}},\ and\
  \bibinfo {author} {\bibfnamefont {Y.}~\bibnamefont {Li}},\ }\bibfield
  {title} {\enquote {\bibinfo {title} {Hydrodynamic behavior of liquid–solid
  micro-fluidized beds determined from bed expansion},}\ }\href@noop {}
  {\bibfield  {journal} {\bibinfo  {journal} {Particuology}\ }\textbf {\bibinfo
  {volume} {38}},\ \bibinfo {pages} {103--112} (\bibinfo {year}
  {2018})}\BibitemShut {NoStop}%
\bibitem [{\citenamefont {Rao}\ \emph {et~al.}(2010)\citenamefont {Rao},
  \citenamefont {Curtis}, \citenamefont {Hancock},\ and\ \citenamefont
  {Wassgren}}]{Rao}%
  \BibitemOpen
  \bibfield  {author} {\bibinfo {author} {\bibfnamefont {A.}~\bibnamefont
  {Rao}}, \bibinfo {author} {\bibfnamefont {J.~S.}\ \bibnamefont {Curtis}},
  \bibinfo {author} {\bibfnamefont {B.~C.}\ \bibnamefont {Hancock}},\ and\
  \bibinfo {author} {\bibfnamefont {C.}~\bibnamefont {Wassgren}},\ }\bibfield
  {title} {\enquote {\bibinfo {title} {The effect of column diameter and bed
  height on minimum fluidization velocity},}\ }\href@noop {} {\bibfield
  {journal} {\bibinfo  {journal} {AIChE J.}\ }\textbf {\bibinfo {volume}
  {56}},\ \bibinfo {pages} {2304--2311} (\bibinfo {year} {2010})}\BibitemShut
  {NoStop}%
\bibitem [{\citenamefont {Wang}\ and\ \citenamefont {Fan}(2011)}]{Wang3}%
  \BibitemOpen
  \bibfield  {author} {\bibinfo {author} {\bibfnamefont {F.}~\bibnamefont
  {Wang}}\ and\ \bibinfo {author} {\bibfnamefont {L.-S.}\ \bibnamefont {Fan}},\
  }\bibfield  {title} {\enquote {\bibinfo {title} {Gas-solid fluidization in
  mini- and micro-channels},}\ }\href@noop {} {\bibfield  {journal} {\bibinfo
  {journal} {Ind. Eng. Chem. Res.}\ }\textbf {\bibinfo {volume} {50}},\
  \bibinfo {pages} {4741--4751} (\bibinfo {year} {2011})}\BibitemShut {NoStop}%
\bibitem [{\citenamefont {Doroodchi}\ \emph {et~al.}(2012)\citenamefont
  {Doroodchi}, \citenamefont {Peng}, \citenamefont {Sathe}, \citenamefont
  {Abbasi-Shavazi},\ and\ \citenamefont {Evans}}]{Doroodchi}%
  \BibitemOpen
  \bibfield  {author} {\bibinfo {author} {\bibfnamefont {E.}~\bibnamefont
  {Doroodchi}}, \bibinfo {author} {\bibfnamefont {Z.}~\bibnamefont {Peng}},
  \bibinfo {author} {\bibfnamefont {M.}~\bibnamefont {Sathe}}, \bibinfo
  {author} {\bibfnamefont {E.}~\bibnamefont {Abbasi-Shavazi}},\ and\ \bibinfo
  {author} {\bibfnamefont {G.~M.}\ \bibnamefont {Evans}},\ }\bibfield  {title}
  {\enquote {\bibinfo {title} {Fluidisation and packed bed behaviour in
  capillary tubes},}\ }\href@noop {} {\bibfield  {journal} {\bibinfo  {journal}
  {Powder Technol.}\ }\textbf {\bibinfo {volume} {223}},\ \bibinfo {pages}
  {131--136} (\bibinfo {year} {2012})}\BibitemShut {NoStop}%
\bibitem [{\citenamefont {C\'u\~nez}\ and\ \citenamefont
  {Franklin}(2023)}]{Cunez5}%
  \BibitemOpen
  \bibfield  {author} {\bibinfo {author} {\bibfnamefont {F.~D.}\ \bibnamefont
  {C\'u\~nez}}\ and\ \bibinfo {author} {\bibfnamefont {E.~M.}\ \bibnamefont
  {Franklin}},\ }\bibfield  {title} {\enquote {\bibinfo {title} {Miniaturized
  gas–solid fluidized beds},}\ }\href
  {https://doi.org/https://doi.org/10.1016/j.mechrescom.2023.104146} {\bibfield
   {journal} {\bibinfo  {journal} {Mech. Res. Commun.}\ }\textbf {\bibinfo
  {volume} {131}},\ \bibinfo {pages} {104146} (\bibinfo {year}
  {2023})}\BibitemShut {NoStop}%
\bibitem [{\citenamefont {Formisani}, \citenamefont {Girimonte},\ and\
  \citenamefont {Longo}(2008)}]{formisani_fluidization_2008}%
  \BibitemOpen
  \bibfield  {author} {\bibinfo {author} {\bibfnamefont {B.}~\bibnamefont
  {Formisani}}, \bibinfo {author} {\bibfnamefont {R.}~\bibnamefont
  {Girimonte}},\ and\ \bibinfo {author} {\bibfnamefont {T.}~\bibnamefont
  {Longo}},\ }\bibfield  {title} {\enquote {\bibinfo {title} {The fluidization
  process of binary mixtures of solids: {Development} of the approach based on
  the fluidization velocity interval},}\ }\href
  {https://doi.org/10.1016/j.powtec.2007.10.003} {\bibfield  {journal}
  {\bibinfo  {journal} {Powder Technol.}\ }\textbf {\bibinfo {volume} {185}},\
  \bibinfo {pages} {97--108} (\bibinfo {year} {2008})}\BibitemShut {NoStop}%
\bibitem [{\citenamefont {Bertsekas}(1992)}]{bertsekas_auction_1992}%
  \BibitemOpen
  \bibfield  {author} {\bibinfo {author} {\bibfnamefont {D.~P.}\ \bibnamefont
  {Bertsekas}},\ }\bibfield  {title} {\enquote {\bibinfo {title} {Auction
  algorithms for network flow problems: {A} tutorial introduction},}\ }\href
  {https://doi.org/10.1007/BF00247653} {\bibfield  {journal} {\bibinfo
  {journal} {Computational Optimization and Applications}\ }\textbf {\bibinfo
  {volume} {1}},\ \bibinfo {pages} {7--66} (\bibinfo {year}
  {1992})}\BibitemShut {NoStop}%
\bibitem [{\citenamefont {Kalman}(1960)}]{Kalman}%
  \BibitemOpen
  \bibfield  {author} {\bibinfo {author} {\bibfnamefont {R.~E.}\ \bibnamefont
  {Kalman}},\ }\bibfield  {title} {\enquote {\bibinfo {title} {A new approach
  to linear filtering and prediction problems},}\ }\href@noop {} {\bibfield
  {journal} {\bibinfo  {journal} {J. Basic Eng. - T. ASME}\ }\textbf {\bibinfo
  {volume} {82}},\ \bibinfo {pages} {35--45} (\bibinfo {year}
  {1960})}\BibitemShut {NoStop}%
\bibitem [{\citenamefont {Del~Pozo}, \citenamefont {Briens},\ and\
  \citenamefont {Wild}(1992)}]{del_pozo_effect_1992}%
  \BibitemOpen
  \bibfield  {author} {\bibinfo {author} {\bibfnamefont {M.}~\bibnamefont
  {Del~Pozo}}, \bibinfo {author} {\bibfnamefont {C.~L.}\ \bibnamefont
  {Briens}},\ and\ \bibinfo {author} {\bibfnamefont {G.}~\bibnamefont {Wild}},\
  }\bibfield  {title} {\enquote {\bibinfo {title} {Effect of column inclination
  on the performance of three-phase fluidized beds},}\ }\href@noop {}
  {\bibfield  {journal} {\bibinfo  {journal} {AIChE Journal}\ }\textbf
  {\bibinfo {volume} {38}},\ \bibinfo {pages} {1206--1212} (\bibinfo {year}
  {1992})}\BibitemShut {NoStop}%
\bibitem [{\citenamefont {O'Dea}\ \emph {et~al.}(1990)\citenamefont {O'Dea},
  \citenamefont {Rudolph}, \citenamefont {Chong},\ and\ \citenamefont
  {Leung}}]{odea_effect_1990}%
  \BibitemOpen
  \bibfield  {author} {\bibinfo {author} {\bibfnamefont {D.~P.}\ \bibnamefont
  {O'Dea}}, \bibinfo {author} {\bibfnamefont {V.}~\bibnamefont {Rudolph}},
  \bibinfo {author} {\bibfnamefont {Y.~O.}\ \bibnamefont {Chong}},\ and\
  \bibinfo {author} {\bibfnamefont {L.~S.}\ \bibnamefont {Leung}},\ }\bibfield
  {title} {\enquote {\bibinfo {title} {The effect of inclination on fluidized
  beds},}\ }\href@noop {} {\bibfield  {journal} {\bibinfo  {journal} {Powder
  Technol.}\ }\textbf {\bibinfo {volume} {63}},\ \bibinfo {pages} {169--178}
  (\bibinfo {year} {1990})}\BibitemShut {NoStop}%
\bibitem [{\citenamefont {Jiang}\ \emph {et~al.}(2018)\citenamefont {Jiang},
  \citenamefont {Hagemeier}, \citenamefont {Bück},\ and\ \citenamefont
  {Tsotsas}}]{jiang_color-ptv_2018}%
  \BibitemOpen
  \bibfield  {author} {\bibinfo {author} {\bibfnamefont {Z.}~\bibnamefont
  {Jiang}}, \bibinfo {author} {\bibfnamefont {T.}~\bibnamefont {Hagemeier}},
  \bibinfo {author} {\bibfnamefont {A.}~\bibnamefont {Bück}},\ and\ \bibinfo
  {author} {\bibfnamefont {E.}~\bibnamefont {Tsotsas}},\ }\bibfield  {title}
  {\enquote {\bibinfo {title} {Color-{PTV} measurement and {CFD}-{DEM}
  simulation of the dynamics of poly-disperse particle systems in a pseudo-{2D}
  fluidized bed},}\ }\href {https://doi.org/10.1016/j.ces.2018.01.013}
  {\bibfield  {journal} {\bibinfo  {journal} {Chem. Eng. Sci.}\ }\textbf
  {\bibinfo {volume} {179}},\ \bibinfo {pages} {115--132} (\bibinfo {year}
  {2018})}\BibitemShut {NoStop}%
\bibitem [{\citenamefont {Oliveira}\ and\ \citenamefont
  {Franklin}(2023)}]{Supplemental3}%
  \BibitemOpen
  \bibfield  {author} {\bibinfo {author} {\bibfnamefont {H.~B.}\ \bibnamefont
  {Oliveira}}\ and\ \bibinfo {author} {\bibfnamefont {E.~M.}\ \bibnamefont
  {Franklin}},\ }\bibfield  {title} {\enquote {\bibinfo {title} {Dataset for
  '{B}idisperse micro fluidized beds: {E}ffect of bed inclination on
  mixing'},}\ }\href {https://doi.org/10.17632/v9c5bsfz7t.1} {\bibfield
  {journal} {\bibinfo  {journal} {Mendeley Data}\ } (\bibinfo {year} {2023}),\
  10.17632/v9c5bsfz7t.1}\BibitemShut {NoStop}%
\end{thebibliography}%

\end{document}